\documentclass[12pt]{article}
\usepackage[utf8]{inputenc}
\usepackage[OT1]{fontenc}
\usepackage{amsmath}
\usepackage{amsthm}
\usepackage{amsfonts}
\usepackage{amssymb}
\usepackage{makeidx}
\usepackage{enumitem}
\usepackage[numbers]{natbib}
\usepackage[T1]{fontenc}
\usepackage{tabularx}
\usepackage[dvips]{epsfig}
\usepackage[dvips]{graphics}
\usepackage{dsfont}
\usepackage{float}
\usepackage{hyperref}
\usepackage{fix-cm}
\usepackage{color}
\usepackage{url}
\hypersetup{
    colorlinks=true, 
    linktoc=all,     
    linkcolor=blue,  
    citecolor=blue,
    filecolor=blue,
    urlcolor=blue,
}

\usepackage{framed}

\newtheorem{t1}{\textbf{Theorem}}

\newtheorem{t3}{\textbf{Remark}}
\newtheorem{t4}{\textbf{Lemma}}

\newtheorem{t6}{\textbf{Proof}}

\begin{document}

\begin{center}
{\Large \bf Penalised inference for autoregressive moving average models with time-dependent predictors}\\[3mm]
Hamed Haselimashhadi and Veronica Vinciotti  \\
Department of Mathematics, Brunel University London, UK
\end{center}
Abstract: Linear models that contain a time-dependent response and explanatory variables have attracted much interest in recent years. The most general form of the existing approaches is of a linear regression model with autoregressive moving average residuals. The addition of the moving average component results in a complex model with a very challenging implementation. In this paper,
we propose to account for the time dependency in the data by explicitly adding autoregressive terms of the response variable in the linear model. In addition, we consider an autoregressive process for the
errors in order to capture complex dynamic relationships parsimoniously. To broaden the application of the model, we present an $l_1$ penalized likelihood approach for the estimation of the parameters and show how the adaptive lasso penalties lead to an estimator which enjoys the oracle property. Furthermore, we prove the consistency of the estimators with respect to the mean squared prediction error in high-dimensional settings, an aspect that has not been considered by the existing time-dependent regression models. A simulation study and real data analysis show the successful applications of the model on financial data on stock indexes.

\noindent Keywords: time series, high dimensional models, lasso

\section{{Introduction}}

This paper deals with fitting a general time series-regression model using $l_1$ regularized inference. In the context of linear models, $l_1$ penalized approaches have received great interest in recent years as they allow to perform variable selection and parameter estimation simultaneously for any data, including high-dimensional datasets, where classical approaches for parameter estimation break down, e.g. \cite{tibshirani1996,knight2000,meinshausen2006,zou2006,park2008}. \cite{zou2006} have shown that a model where penalties are adapted to each individual regressor enjoys oracle properties. Most of the advances in regularized regression models have been for the case of independent and identically distributed data. A recent line of research has concentrated on regularized models in time dependent frameworks. Amongst these, \cite{wang2007} showed the successful application of $l_1$ penalised inference  in the context of  autocorrelated residuals for a fixed order, by proposing the model
\begin{equation*}
y_t=\sum_{i=1}^{\text{r}}x'_{ti}\beta_i+\sum_{j=1}^{q}\theta_j\epsilon_{t-j}+e_t,
\end{equation*}
and studied the properties of this model in low-dimensional settings. \cite{nardi2010} studied the theoretical properties of a regularized autoregressive process on $Y_t$ for both low and dimensional cases, whereas \cite{song2011} studied the $l_1$ estimation of vector autoregressive models. In both cases, no exogenous variables are included in the model. \cite{marcelo2012} studied the asymptotic properties of adaptive lasso in high dimensional time series models when the number of variables increases as a function of the number of observations. Their model covers a lagged regression in the presence of exogenous variables, but does not consider autocorrelated residuals. Recently, \cite{wu2012shrinkage} proposed an extension of the model of \cite{wang2007}  by adding a moving average term, that is they propose a model of the form
\begin{align*}
& y_t=\sum_{i=1}^{\text{r}}x'_{ti}\beta_i+\epsilon_t, \qquad
\epsilon_t=\sum_{j=1}^{q}\theta_j\epsilon_{t-j}+e_t+\sum_{j=1}^{q}\phi_j e_{t-j}.
\end{align*}
Similarly to \cite{wang2007}, they proved the consistency of the model in low-dimensional cases. Despite the generality of this model, considering an ARMA process for the errors results in a complex model with a challenging implementation.

\begin{figure}[H]
	\centering
	\includegraphics[width=.5\linewidth,keepaspectratio=true]{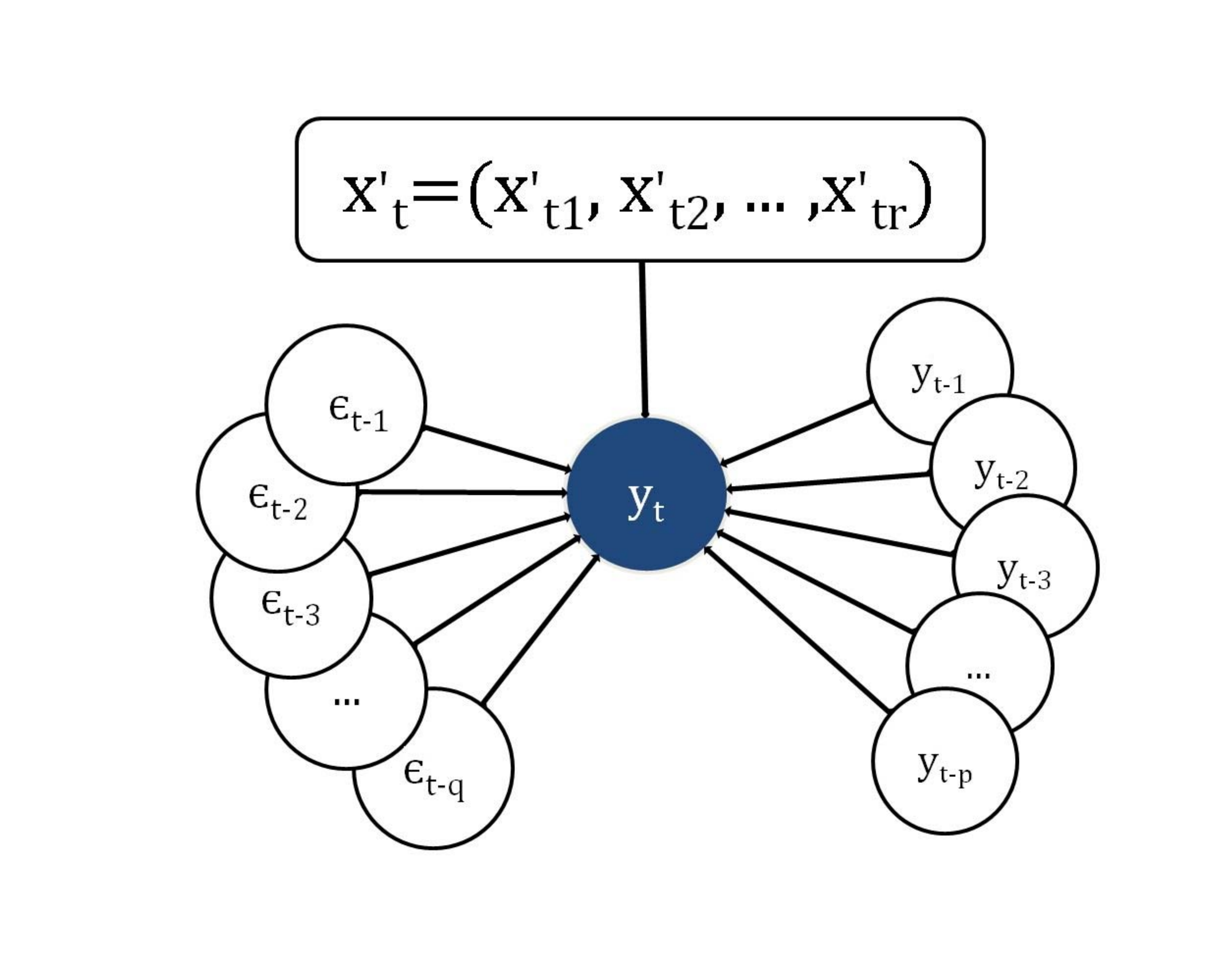}
	\caption{Schematic illustration of a REGARMA(p,q) model}
	\label{figure:1}
\end{figure}

In this paper, we propose to account for the time dependency in the data by explicitly adding autoregressive terms of the response variable in the linear model, as in \cite{nardi2010}, as well as an autocorrelated process for residuals, as in \cite{wang2007}, in order to capture complex dynamics parsimoniously. In particular, given fixed orders $p$ and $q$, we propose the model
\begin{equation}\label{REGARMA}
	y_t=x'_t\beta+\sum_{j=1}^{p}\phi_jy_{t-j}+\sum_{i=1}^{q}\theta_i \epsilon_{t-i}+e_t.
\end{equation}
We name the terms in the right hand side of (\ref{REGARMA}) as \textbf{REG}ression term, \textbf{A}uto\textbf{R}egressive term and \textbf{M}oving \textbf{A}verage term respectively and call the resulting model REGARMA. 
We assume that all time dependent components in REGARMA are stationary and ergodic. Figure (\ref{figure:1}) illustrates a schematic view of the model. 
 In Section \ref{Definitions}, we formulate the model and present an $l_1$ penalized likelihood approach for the estimation of the parameters. In Section \ref{TheoreticalPropertiesOfREGARMA}, we prove the asymptotic properties of the model and show how the adaptive lasso penalty leads to estimators which enjoy the oracle property. Furthermore, we prove  the consistency of the model with respect to the mean squared prediction error in high-dimensional settings, an aspect that has not been considered by the existing time-dependent regression models.  In section \ref{Algorithm}, we discuss the implementation of REGARMA. A simulation study, given in section \ref{simulation studi}, will accompany the theoretical results. In section \ref{RealData} we apply the model to two real datasets in finance and macroeconomic, respectively. Finally, we draw some conclusions in section \ref{Conclusion and discussion}.

\section{{$L_1$ penalised parameter estimation of REGARMA}}\label{Definitions}
\normalsize
The general form of REGARMA consists of a lagged response variable, covariates and autocorrelated residuals.
Consider the following Gaussian REGARMA model of order $p$ and $q$,
\begin{equation*}
		y_t=x'_t\beta+\sum_{j=1}^{p}\phi_jy_{t-j}+\sum_{i=1}^{q}\theta_i \epsilon_{t-i}+e_t, \qquad e^{}_t \overset{iid}{\sim} N(0,\sigma_{}^2),  \qquad t=1,2,3, \ldots ,T
\end{equation*}
where  $x'_t$ is the $t^{th}$ row of the matrix of $r$ predictors $X'_{T\times r}$, $\{y_t\}$ and $\{\epsilon_t\}$
follow stationary time series processes, that is all roots of the polynomials $1-\sum_{i=0}^{p}\phi_i L^i$ and $1-\sum_{i=0}^{q}\theta_i L^i$ are unequal and outside the unit circle, $e^{}_t$s are independent and identical Gaussian noises with mean of zero and  finite fourth moments, 
and $p$ and $q$ are both less than the number of observations $T$. Moreover, we assume that the errors and explanatory variables in $X$ are independent of each other.
To remove the constants from the model we follow the literature on regularized models, e.g. \cite{tibshirani1996,huang2008}, and standardize the covariates and response to zero means and unit variance.

Given the first $T_\circ=p+q$ observations, maximizing the $l_1$ penalized conditional likelihood of the model is equivalent to minimizing
\small
\begin{equation}\label{openmodel1}
	\begin{aligned}
		Q_n(\Theta)= & \sum_{t=T_\circ+1}^{T}\left((y_t-x'_t\beta)-\sum_{i=1}^{p}\phi_iy_{t-i}-\sum_{j=1}^{q}\theta_j\epsilon_{t-j} \right)^2
		+\sum_{i=1}^{r}\lambda|\beta_i|
		+\sum_{j=1}^{p}\gamma|\phi_j|
		+\sum_{k=1}^{q}\tau|\theta_k|\\
	\end{aligned}
\end{equation}
\normalsize
where $\lambda,\gamma,\tau$  are tuning parameters and $\Theta=(\beta',\phi',\theta')$ is the vector of regression, autoregressive and moving average parameters. Following the literature, and given the superior properties of adaptive lasso models \citep{zou2006}, we also propose an adaptive version of REGARMA penalised estimation as follows
\small
\begin{equation*}\label{modifiedmodel1}
	\begin{aligned}
		Q^*_n(\Theta)= & \sum_{t=T_\circ+1}^{T}\left((y_t-x'_t\beta)-\sum_{i=1}^{p}\phi_iy_{t-i}-\sum_{j=1}^{q}\theta_j\epsilon_{t-j} \right)^2 +\sum_{i=1}^{r}\lambda^*_i|\beta_i|
		+\sum_{j=1}^{p}\gamma^*_j|\phi_j|
		+\sum_{k=1}^{q}\tau^*_k|\theta_k|\\
	\end{aligned}
\end{equation*}
\normalsize
where
$\lambda^*_i,\gamma^*_j,\tau^*_k, i=1,2,\ldots, r;  j=1,2,\ldots, p; k=1,2,\ldots, q$  are  tuning parameters.
\subsection{{Matrix representation of the model}}
For convenience, we write the model in matrix representation.
Let $H'=(H_{(p)},H_{(q)},X')$ be a $n\times(p+q+r)$ matrix including  lags of autoregressive ($H_{(p)}$), moving average ($H_{(q)}$), and explanatory variables ($X'$). Let  $\Theta=(\phi',\theta',\beta')$ denote the vector of corresponding parameters, $e'=(e_{T_0+1},e_2, \ldots ,e_T)$ be the vector of errors, $T_\circ=p+q$ and $n=T-T_\circ$, as previously defined.
Then, in matrix form, the model can be written as
\begin{equation*}\label{matrixform}
Y=H'\Theta+e
\end{equation*}
and the $l_1$ penalized conditional likelihood given the first $T_0$ observation is equivalent to
\begin{equation*}
	Q_n(\Theta)=L(\Theta)+\lambda'|\beta|+\gamma'|\phi|+\tau'|\theta|,
\end{equation*}
where $L(\Theta)=e'e, \lambda'=\{\lambda\}_{1 \times r}, \gamma'=\{\gamma\}_{1 \times p},  \tau'=\{\tau\}_{1 \times q}$. Similarly, the adaptive form of the model is given by
\begin{align}\label{adaptive-REGARMA in matrix}
	Q^*_n(\Theta)=L(\Theta)+\lambda'^*|\beta|+\gamma'^*|\phi|+\tau'^*|\theta|,
\end{align}
where the parameters are given by
\begin{align*}
		\lambda^{*'}  =(\lambda^*_1,\lambda^*_2, \ldots, \lambda^*_{r} ),
		\gamma^{*'} =(\gamma^*_1,\gamma^*_2, \ldots, \gamma^*_p ),
		\tau^{*'}   =(\tau^*_1,\tau^*_2, \ldots, \tau^*_q ),
		\Theta=(\beta',\phi',\theta').
\end{align*}
\section{{Theoretical properties of REGARMA and adaptive-REGARMA}}\label{TheoreticalPropertiesOfREGARMA}
\normalsize
In order to study the theoretical properties of REGARMA and adaptive-REGARMA, we define the true coefficients by $\Theta^\circ=(\beta^{\circ^{'}},\phi^{\circ^{'}},\theta^{\circ^{'}})$
and assume that some of  these coefficients are zero. The indexes of non-zero coefficients in each group of coefficients, $\beta,\phi$ and $\theta$, are denoted by $s_1,s_2$ and $s_3$ respectively, whereas $s^c_1,s^c_2,s^c_3 $ are the complementary sets and contain the indexes of zero coefficients. We also define $\beta^\circ_{s_1},\phi^\circ_{s_2},\theta^\circ_{s_3}$ and their corresponding (REGARMA) estimations by $\hat\beta_{s_1},\hat\phi_{s_2},\hat\theta_{s_3}$. Similarly, adaptive-REGARMA estimations are denoted by $\hat\beta^*_{s_1},\hat\phi^*_{s_2},\hat\theta^*_{s_3}$. Finally, different combinations of model parameters are going to be used, with obvious meaning, in particular
$\Theta_1^\circ=\{\beta^{\circ '}_{s_1},\phi^{\circ'}_{s_2} ,\theta^{\circ'}_{s_3}\}$ ,
$\Theta_2^\circ=\{\beta^{\circ'}_{s^c_1},\phi^{\circ'}_{s^c_2} ,\theta^{\circ'}_{s^c_3}\}$, $\hat \Theta_1=\{\hat \beta^{'}_{s_1},\hat \phi^{'}_{s_2} ,\hat \theta^{'}_{s_3}\}$ ,
$\hat \Theta_2=\{\hat \beta^{'}_{s^c_1},\hat \phi^{'}_{s^c_2} ,\hat \theta^{'}_{s^c_3}\}$,
$ \hat\Theta_1^*=\{\hat \beta^{*'}_{s_1},\hat \phi^{*'}_{s_2} ,\hat \theta^{*'}_{s_3}\}$ ,
$\hat \Theta_2^*=\{\hat \beta^{*'}_{s^c_1},\hat \phi^{*'}_{s^c_2} ,\hat \theta^{*'}_{s^c_3}\}$
.
\subsection{{Assumptions}}\label{Assumptions}
\normalsize
To prove the theoretical properties of the estimators, in line with  the literature, we make use of the following assumptions: 
\begin{enumerate}[label=($\alph*$).,leftmargin=.8in]
	\item  $e_t$s are i.i.d Gaussian random variables with finite fourth moments
	\item  The covariates, $X_i,i=1,2,3,\ldots,r$, and response variable, $Y$, are \textit{stationary} and \textit{ergodic} with finite second order moments. Also, we assume that none of the roots of $1-\sum_{i=1}^{p}\phi_i L^i$ and/or $1-\sum_{j=1}^{q}\theta_j L^i$ are equal and outside of the unit circle
	 \item  $X_i,i=1,2,3,\ldots,r$ are independent of the errors
	 \item  $\frac{1}{n}X'X \rightarrow_{a.s} \mathbb{E}(X'X)<\infty$ and $\max_{1 \leq i \leq r} X_i'X_i<\infty$.
\end{enumerate}
\def \Nassumption {\text{$a-d$}}
Assumptions $(a)-(b)$ are standard assumptions for dealing with  stationary time series.  Assumption $(c)-(d)$ are used to guarantee that explanatory variables have finite expectations.
\subsection{{Theoretical properties of REGARMA} when $r<n$}
In the following theorems, we extend the theorems of \cite{wang2007} to cover a model with a lagged response.
\begin{t1} \label{theorem2}
Assume $\lambda_n\sqrt{n} \rightarrow \lambda_\circ$ , $\gamma_n \sqrt{n} \rightarrow\gamma_\circ$ ,  $\tau_n \sqrt{n} \rightarrow \tau_\circ$ and $\lambda_\circ,\gamma_\circ,\tau_\circ \geq 0$. Then under assumptions \Nassumption, it follows that $\sqrt{n}(\hat\Theta-\Theta^\circ ) \overset{d}{\rightarrow} \arg\min{(k(\delta))} $
	where
	\begin{align*}
			k(\delta)=  -2\delta'W+\delta'U_B\delta
			           & +\lambda_\circ\sum_{i=1}^{r}\{u_i sign(\beta^\circ _i)I(\beta^\circ _i\neq 0)+|u_i|I(\beta_i^\circ =0)  \} \\
			           & +\gamma_\circ\sum_{j=1}^{p}\{v_j sign(\phi^\circ _j)I(\phi^\circ _j\neq 0)+|v_i|I(\phi_j^\circ =0)  \}     \\
			           & +\tau_\circ\sum_{k=1}^{q}\{w_k sign(\theta^\circ _k)I(\theta^\circ _k\neq 0)+|w_k|I(\theta_k^\circ =0)  \}
	\end{align*}
with $\delta=(u',v',w')$ is a vector of parameters in $\mathbb{R}^{(r+p+q)}$, $W \sim MVN(O,\sigma^2U_B)$
and $U_B=\mathbb{E}(HH')$.
\end{t1}
The proof is given in the Appendix. Theorem (\ref{theorem2}) shows that the REGARMA estimator has a Knight-Fu type asymptotic property \citep{knight2000} and it implies that the tuning parameters in $Q_n(\Theta)$ cannot shrink to zero at a speed faster than $n^{-1/2}$. Otherwise, $\{\lambda_\circ,\gamma_\circ,\tau_\circ\}$ are zero and $k(\delta)$ becomes a standard quadratic function,
\begin{equation*}
	 k(\delta)=-2\delta W+\delta' U_B \delta
\end{equation*}
which does not produce a sparse solution. In addition, the proof of theorem (\ref{theorem2}) requires the errors to be independent and identically distributed but we do not make a strong assumption on the type of distribution for the errors, due to the use of the martingale central limit theorem for large $n$. 

\cite{knight2000} proves that a lasso optimization returns estimates of non-zero parameters that suffer an asymptotic bias. This applies also to the REGARMA model, as we show with the following remark.
\begin{t3}\label{lasso bias}
	Consider a special case of REGARMA when $\beta^\circ_i>0, \quad 1 \leq i\leq r$ but $\theta^\circ _{j_1}=0$ and $\phi^\circ _{j_2}=0$ for $1 \leq {j_1 }\leq q ,\, 1\leq j_2 \leq p,\,j_1,\,j_2 \in \mathbb{N}$. If minimizing $ k(\delta)$ can correctly identify $\Theta$, it means that $u \neq 0$ and $v,w=0$. That is, $ k(\delta)$ must satisfy
	\begin{equation*}
		\begin{aligned}
			\frac{\partial  k(\delta)}{\partial u}
			& =\frac{\partial  k(u,0,0)}{\partial{u}}                                                                                                           \\
			& =\frac{\partial}{\partial{u} } \Big(-2(u',0,0)W+(u',0,0)'U_{B}(u',0,0)
			 +(n\lambda'_n|\beta^\circ +\frac{u}{\sqrt{n}}|-n\lambda'_n|\beta^\circ |)
			\Big)                         \\
			& =-2W_{{1:r}}+2u'U_{{B_{1:r}}}+\lambda_\circ1_{r\times 1}=0                                                                                               \\
			  & \Longrightarrow u'=\frac{1}{2}(2W_{{1:r}}-\lambda_\circ 1_{r\times 1})U_{{B_{1:r}}}^{-1}.                                                             
		\end{aligned}
	\end{equation*}
Then using Theorem \ref{theorem2}, $\sqrt{n}(\hat\beta-\beta^\circ) \overset{d}{\rightarrow}\arg\min( k(\delta=u'))=MVN\bigg(\mathbb{E}(u')\neq 0,\, U_{{B_{1:r}}}^{-1}\bigg),$
where $U_{B_{1:r}}^{-1}$ is the matrix with the first $r$ rows of $U_B$ corresponding to the $r$ covariates.
\end{t3}
If $\lambda_\circ,\gamma_\circ$ and $\tau_\circ$ are positive, remark (\ref{lasso bias}) shows that $Q_n(\Theta)$ suffers an asymptotic bias and is different from the oracle estimator, $MVN\Big(O,U^{-1}_{{B_{1:r}}}\Big)$. In other words, REGARMA is not asymptotically consistent unless $\lambda_\circ,\gamma_\circ,\tau_\circ \underset{n\rightarrow \infty}{\rightarrow} 0$. The following remark can be extended to other groups of coefficients.

\subsection{{Theoretical properties of adaptive-REGARMA when $r<n$}}
\normalsize
Following the notation of section \ref{Definitions}, we consider the adaptive version of the penalised likelihood and estimate the model parameters by minimizing
\begin{align*}
	Q^*_n(\Theta)=L_n(\Theta)+n\lambda'^*|\beta|+n\gamma'^*|\phi|+n\tau'^*|\theta|
\end{align*}
where
\begin{align*}
	\begin{array}{l}
		L_n(\Theta)=
		\bigg(
		Y-X'\beta-H_{(p)}\phi+H_{(q)}\theta
		\bigg)'
		\bigg(
		Y-X'\beta-H_{(p)}\phi+H_{(q)}\theta
		\bigg)\\
		\lambda^{*'}=\{\lambda^*\}_{r \times 1}',\,
		\gamma^{*'}=\{\gamma^*\}_{p \times 1}',\,
		\tau^{*'}=\{\tau^*\}_{q \times 1}',\,\Theta=(\beta',\phi',\theta').                                                                                         \\
	\end{array}
\end{align*}
Following \cite{wang2007} and \cite{fan2001}, we define the maximum and minimum penalties for significant and insignificant coefficients by
\begin{align*}
	& a_n=\max(\lambda^*_{i_1},\gamma^*_{i_2},\tau^*_{i_3}; \quad i_1 \in s_1, i_2 \in s_2, i_3 \in s_3)  \label{an}, \\
	& b_n=min(\lambda^*_{i^c_1},\gamma^*_{i^c_2},\tau^*_{i^c_3}; \quad i^c_1 \in s^c_1, i^c_2 \in s^c_2, i^c_3 \in s^c_3), 
\end{align*}
and prove a number of results on the theoretical properties of adaptive REGARMA.
\begin{t1}\label{theorem3}
	Assume $a_n = o(1)$ as $n \rightarrow \infty$. Then under assumptions \Nassumption, there is a local minimiser $\hat \Theta^*$ of $Q_n^*(\Theta)$ such that
	\begin{align*}
		(\hat \Theta^*-\Theta^\circ) = O_p(n^{-1/2}+a_n).
	\end{align*}
\end{t1}
The proof of the theorem is in the Appendix. Let $\alpha_n=a_n+n^{-1/2}$, then, theorem (\ref{theorem3}) proves that there exists a $\sqrt{n}-consistent$ local minimiser $Q_n^*(\Theta)$, when the tuning parameters (for significant variables) of REGARMA converge to zero at the speed {faster} than $n^{-1/2}$ (since $n\alpha^2_n \rightarrow o(1)$).

 As the next step, we prove that if the tuning parameter associated with insignificant variables in REGARMA shrink to zero at a speed {slower} than ${n^{-1/2}}$, then their associated REGARMA coefficients will be estimated exactly equal to zero with probability tending to 1.
\begin{t1}\label{theorem4}
	Assume $b_n\sqrt n \rightarrow \infty$ and $||\hat \Theta^* -\Theta^\circ ||=O_p(n^{-1/2})$ then
	\begin{align*}
		\begin{aligned}
			Pr(\hat\beta^*_{s_1^c}=0)  & \rightarrow 1,  \quad
			Pr(\hat\phi^*_{s_2^c}=0)   & \rightarrow 1, \quad
			Pr(\hat\theta^*_{s_3^c}=0) & \rightarrow 1.
		\end{aligned}
	\end{align*}
\end{t1}
The proof of the theorem is in the Appendix. 
Theorem (\ref{theorem3}) and (\ref{theorem4}) indicate that $\sqrt{n}-consistent$ estimator $\hat{\Theta}^*$ satisfies $Pr(\hat{\Theta}^*_2=0) \rightarrow 1$ under certain conditions on the tuning parameters, leading to the following result:
\begin{t1}\label{theorem5}
	Assume $a_n\sqrt{n} \rightarrow 0$ and $b_n\sqrt{n} \rightarrow \infty$. Then, under assumptions \Nassumption, the component $\hat\Theta_1^*$ of the local minimiser of $\hat\Theta^*$ in Theorem 3 satisfies
	\begin{align*}
		\sqrt{n}(\hat \Theta^*_1-\Theta^\circ _1) \overset{d}{\rightarrow}MVN(O,\sigma^2U_0^{-1})
	\end{align*}
	where $U_0$ is the sub-matrix $U_B$ corresponding to $\Theta_1^\circ $.
\end{t1}
The proof of the theorem is in the Appendix. Theorem (\ref{theorem5}) implies that if $a_n$ tends to zero at the speed faster than $\sqrt{n}$ 
and simultaneously $b_n$ increases at the speed slower than $\sqrt{n}$, 
then adaptive REGARMA is asymptotically an oracle estimator. In the next subsection, we consider the theoretical properties of adaptive REGARMA for high-dimensional problems.

\subsection{{Theoretical properties of adaptive REGARMA when $n \ll r$}}
In the proofs of the low-dimensional results (refer to proof of theorem \ref{theorem2}), we rely on a unique path of reaching the maximum of the log likelihood. This is not true in high-dimensional cases, so different results are needed in this case. In this section we follow a similar strategy to~\cite{sourvan2013} to prove theorems in the high dimensional case, an aspect which has not been considered by existing time-dependent regression models, such as those of \cite{wang2007} and \cite{wu2012shrinkage}.

In order to study the consistency of REGARMA in high-dimensional situations, we show that under assumptions \Nassumption, REGARMA is consistent with respect to the mean squared prediction error.\\
 Without loss of generality, we define the REGARMA model as a constrained optimization \citep{tibshirani1996}. Thus, we have
\begin{equation}\label{REGARMA:original:form:in Tibshirani}
\begin{aligned}
&\min\{(y-  X'\beta-H_{(p)}\phi-H_{(q)}\theta)'(y-X'\beta-H_{(p)}\phi-H_{(q)}\theta)\} \\
& \text{ Subject to }
\sum_{j=1}^{r}|\beta_j|\leq K_\lambda, \quad
\sum_{k=1}^{p}|\phi_k| \leq K_\gamma, \quad
\sum_{l=1}^{q}|\theta_l|\leq K_\tau, \\
& and  \hspace{30pt} K_\lambda\geq 0, \quad K_\gamma \geq 0, \quad K_\tau \geq 0
\end{aligned}
\end{equation}
\normalsize
where there is  a one-to-one correspondence between $\lambda,\gamma$ and $\tau$ in REGARMA, and $K_\lambda,K_\gamma$ and $K_\tau$ in (\ref{REGARMA:original:form:in Tibshirani}).
Define the \textit{Mean Squared Prediction Error}, $(MSPE)$, and its estimated value, $\widehat{MSPE}$, by
\begin{align*}
 & MSPE(\hat\beta,\hat\phi,\hat\theta)=\mathbb{E}(\Vert  \hat{Y} -Y^\circ \Vert ^2), \quad
 & \widehat{ MSPE}(\hat\beta,\hat\phi,\hat\theta)=\frac{1}{n}\Vert  \hat{Y} -Y^\circ \Vert^2
\end{align*}
where $Y^\circ$ and $ \hat{Y} $ are the REGARMA predictions of $Y$ based on the true parameters $(\beta^\circ,\phi^\circ,\theta^\circ)$ and REGARMA estimates $(\hat\beta,\hat\phi,\hat\theta)$  from (\ref{REGARMA:original:form:in Tibshirani}), respectively. Then the following theorem holds.
\begin{t1}\label{theorem:Consistency}
Under assumptions \Nassumption\, and $\Vert X \Vert_\infty \leq  M_1, \Vert H_{(p)} \Vert_\infty \leq M_2$, $\Vert {H_{(q)}} \Vert_\infty \leq M_3$ and $M_{max}=\sup\{M_1,M_2,M_3\}$, let $\hat\beta^{},\hat\phi^{}$ and $\hat\theta^{}$ be the REGARMA estimates, and $K_{max}=
\sup
\{
K_\lambda,K_\gamma,K_\tau
\}$ such that
\begin{align*}
\sum_{j=1}^{r}|\beta_j|\leq K_\lambda< \infty, \quad
\sum_{k=1}^{p}|\phi_k|\leq K_\gamma < \infty, \quad
\sum_{l=1}^{q}|\theta_l|\leq K_\tau < \infty.
\end{align*}
Then
\fontsize{10pt}{10pt}
\begin{align}\label{consistency:eq1}
\widehat{ MSPE}(\hat\beta^{},\hat\phi^{},\hat\theta^{})
 \leq
 \frac{ 2K_{max} M_{max}\sigma}{\sqrt{n}}
 \left(\sqrt{2\log(2r)}+
\sqrt{2\log(2p)}+
\sqrt{2\log(2q)}
\right).
\end{align}
\normalsize
\end{t1}
The proof of the theorem is in the Appendix. Note that in the situation where $p=0$ and $q=0$, equation (\ref{consistency:eq1}) results in the standard lasso consistency formula in~\cite{sourvan2013}. When $K_{max}$ is correctly chosen, equation (\ref{consistency:eq1}) also shows that REGARMA is prediction consistent when $\max\{\log(r) ,\log(p) ,\log(q)\} \ll n $.\\
It is also possible to extend this result to  $MSPE$.
\begin{t3}\label{remark 2}
Under the same conditions as Theorem (\ref{theorem:Consistency}),
\small
\begin{align*}
 MSPE(\hat\beta^{},\hat\phi^{},\hat\theta^{})
 \leq
 \frac{ 2K_{max} M_{max}\sigma}{\sqrt{n}}
 \sum_{i=1}^{3}
 \left(\sqrt{2\log(2a_i)}
\right)+
8K^*\sum_{i,j=1}^{3}
\left(
M_i M_j\sqrt{\frac{2log(2 a_ia_j)}{n}}
\right),
\end{align*}
\normalsize
where $K_{\rm max}$ and $M_{max}$ are defined as before, $K^*$ is defined in the Appendix and $a_1=r, a_2=p$ and $a_3=q$.
\end{t3}
Remark (\ref{remark 2}) shows that if $\underset{i,j=1,2,3 }{max}\log(a_ia_j) \ll n$ then REGARMA is consistent.  Given that relatively small orders $p$ and $q$ are sufficient for most time series analyses, the consistency of the estimator is mainly dominated by the high-dimensional regression part. If $M_1 \geq M_2 \geq M_3$ then the above equation approximately reduces to a form similar to the standard lasso results in~\cite{sourvan2013},
\begin{align*} 
 MSPE(\hat\beta^{},\hat\phi^{},\hat\theta^{})
 \leq
 \frac{ 2K_\lambda M_1\sigma}{\sqrt{n}}
 \sqrt{2\log(2r)}
+
8K^*
M_1^2\sqrt{\frac{2log(2 r^2)}{n}}
.
\end{align*}
\normalsize
\section{{Algorithm}}\label{Algorithm}
Since $Q_n(\Theta) \subseteq  Q^*_n(\theta)$, that is REGARMA is a subset of adaptive-REGARMA, and given the improved properties of adaptive REGARMA, we mainly focus on adaptive REGARMA in this section. Our formulation of the model lends itself naturally to its implementation, in contrast to the more complex implementation of the model of \cite{wu2012shrinkage}.

As the model contains regressions, moving averages and autoregressive coefficients, we use the two-step optimization procedure
\begin{align}\label{Alrorithem:eq1}
\begin{aligned}
	\text{First step: }  \hat\epsilon=Y-X'\hat\beta-H_{(p)}\hat\phi, \quad
	\text{Second step: }  Y=X'\beta+H_{(p)}\phi+\hat H_{(q)}\theta. \nonumber
\end{aligned}
\end{align}
Steps 1 and 2 provide a solution to REGARMA  using the adaptive-Lasso algorithm of \cite{zou2006}.

In terms of the selection of the penalties $\lambda$, $\gamma$ and $\tau$, these can be chosen using K-fold cross-validation 
or using an information criterion such as BIC or AIC, CP similarly to \cite{wang2007},~\cite{wu2012shrinkage} and~\citep{hirose2011efficient}. The weights in adaptive-REGARMA are defined by using the (non-adaptive) REGARMA estimates.
Some notes are needed about the selection of the orders  $p$ and $q$ in the REGARMA model. We propose two general approaches to choose the optimal orders for the model: (a) setting an upper bound $P$ and $Q$ and choosing the model that minimizes BIC or AIC inside these bounds (b)\label{method b page number} setting an upper bound $P$ and $Q$ and letting the model choose the best orders by keeping or eliminating the time series coefficients under $L_1$ sparsity constraints. These two approaches are very similar but there is a slight difference between them: in the second approach, the fitting is based on $n=T-(P+Q)$ time points, whereas in the first approach, the number of time points depends on the orders p and q. Then a rule of thumb is to use the first approach when the number of observations is low and choose the second approach when there are enough observations. 

We are in the process of implementing the methods into an R package. This is particularly needed in this area as, to the best of our knowledge, there is no implementation available. The current version of the package is available at \url{http://people.brunel.ac.uk/~mastvvv/Software/}.

\section{{Simulation study}}\label{simulation studi}
We design a simulation study to compare the REGARMA model with existing methods. In the simulation, we:

\begin{enumerate}
\item Set the proportion of zero coefficients to 90\%, 50\% or 10\%.
\item Assign unequal random numbers in  $(-1,1)$ to each non-zero coefficient.
\item Generate the design matrix, $X$, using stationary Gaussian processes, with $r=25,75,200,300,400$ and $T=50,100,150,200,250$.
\item Generate  $ e \sim \sigma \times N(0,1)$ where $\sigma \in \{0.5,1,1.5\}$,
\item Set unequal AR and MA parameters, under the constraint that the roots of stationary polynomials site outside the unit circle  and simulate data from a REGARMA model with $p \leq 3$ and $q\leq 3$.
\item Repeat each combination of models 10 times.
\end{enumerate}
We compare the adaptive REGARMA model with adaptive lasso, as it is the closest model in the literature for which an implementation is available. Similar results were found in the comparison of the non-adaptive versions (results not reported). BIC was used to choose the optimal penalties, whereas the autoregressive and moving average orders were fixed as the true ones.

\begin{figure}[th!]
 	\centering
     \includegraphics[scale=.54]{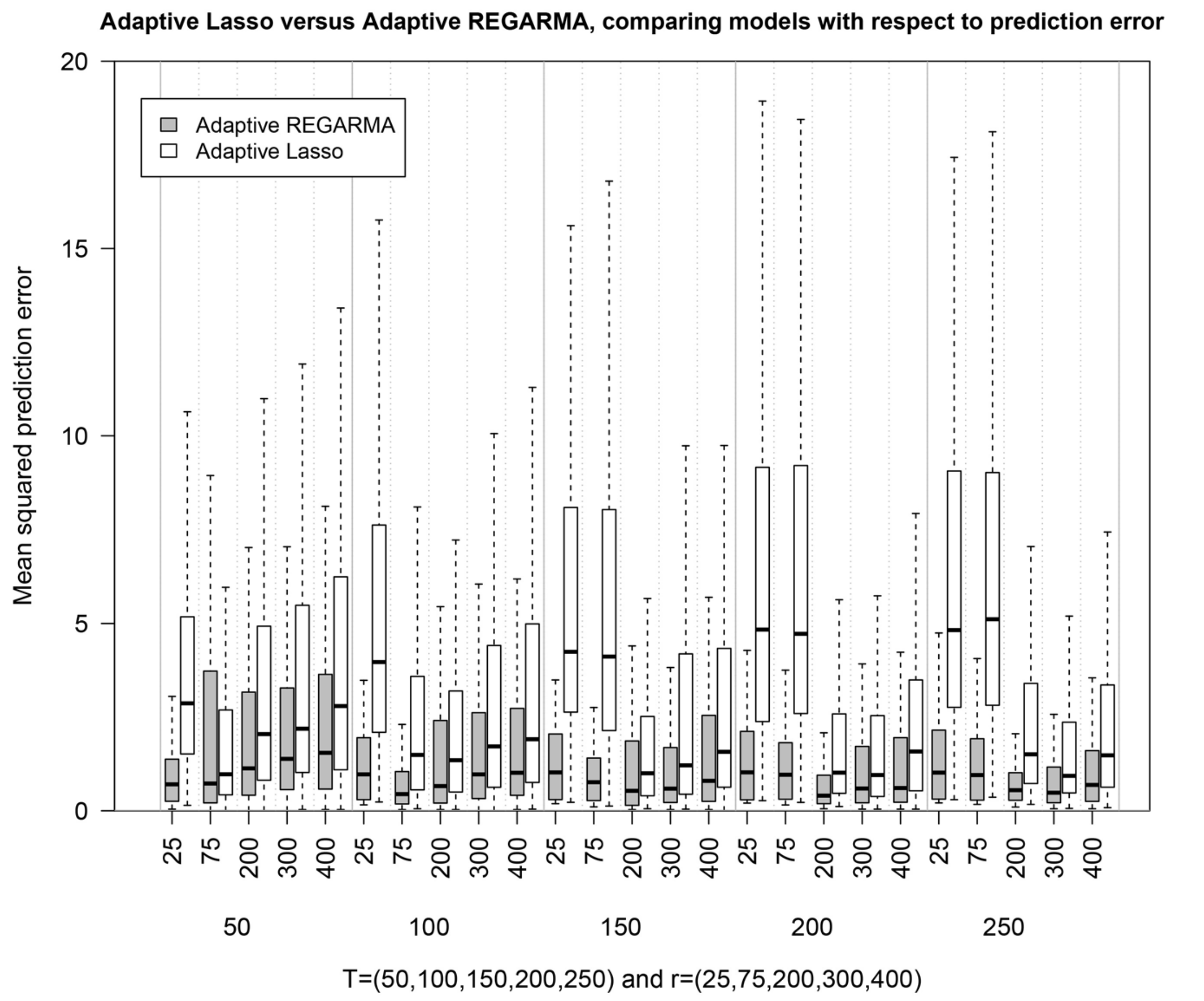}
 	\caption{Comparison of adaptive lasso and adaptive REGARMA with respect to mean squared prediction error on simulated data with different values of $r$ and $T$.}
 	\label{figure:sim:1}
 \end{figure}

\begin{figure}[th!]
 	\centering
     \includegraphics[scale=.50]{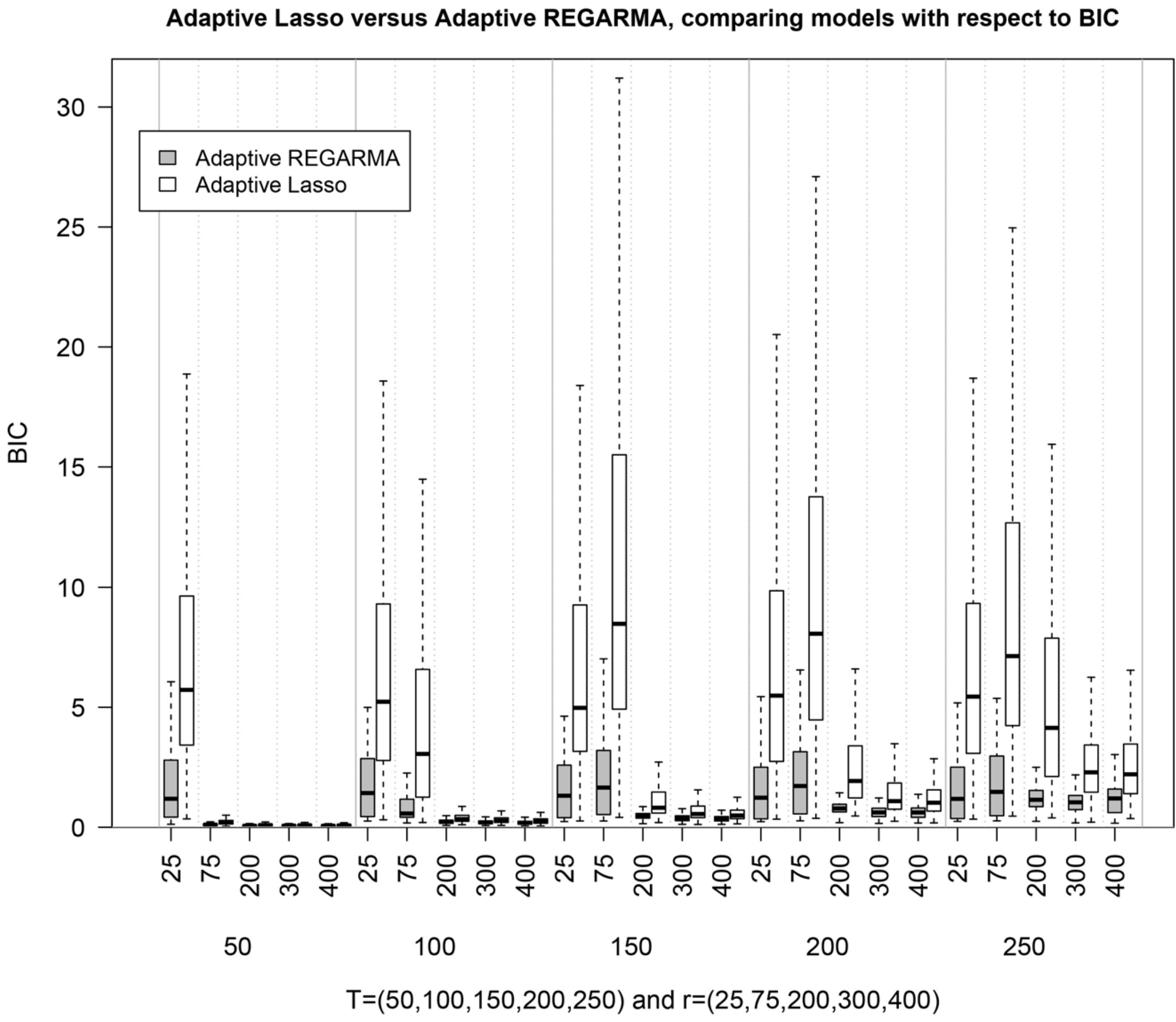}
 	\caption{Comparison of adaptive lasso and adaptive REGARMA with respect to BIC on simulated data with different values of $r$ and $T$.}
 	\label{figure:sim:2}
 \end{figure}

\begin{figure}[th!]
	\centering
    \includegraphics[scale=.50]{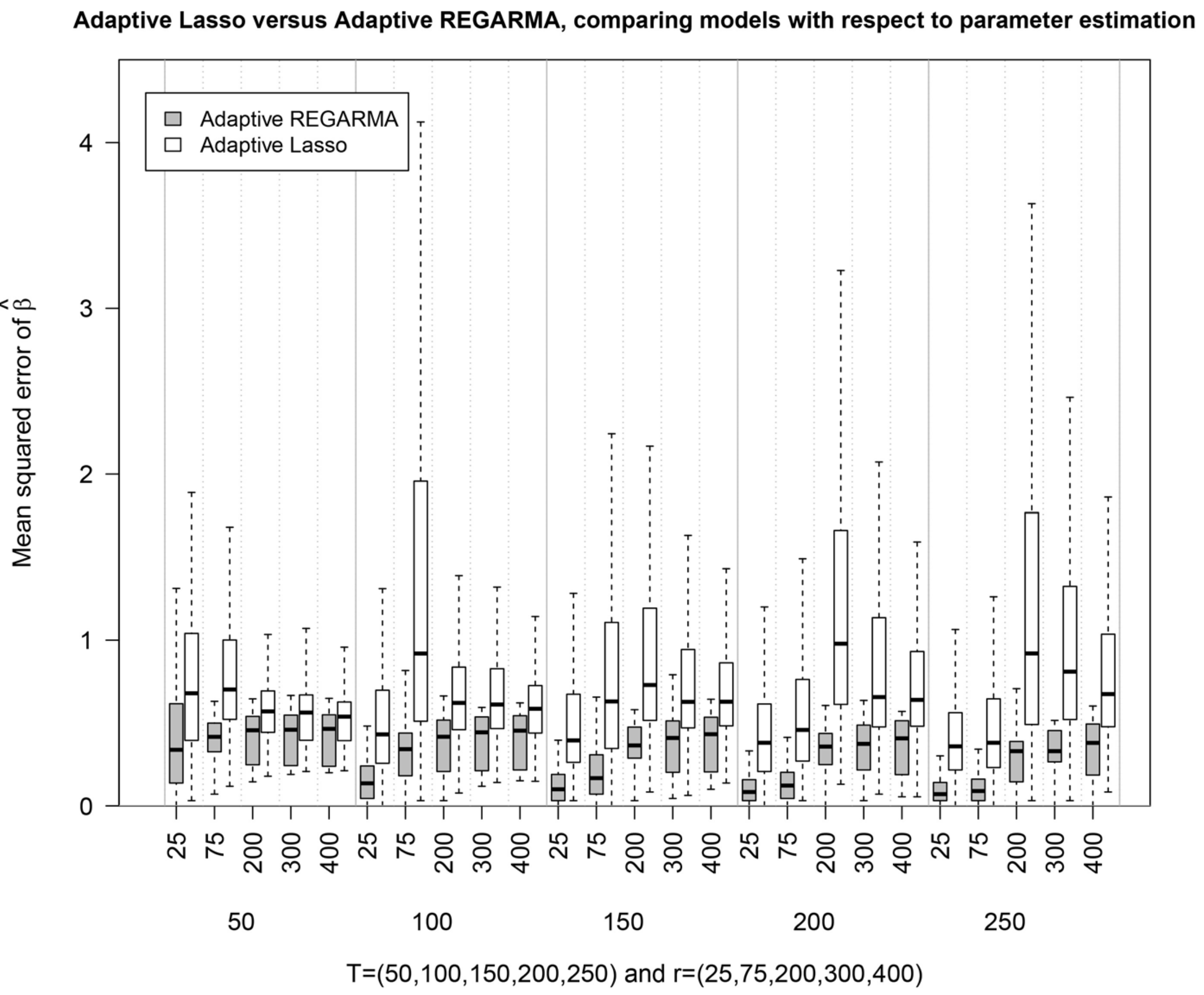}
	\caption{Comparison of adaptive Lasso and adaptive REGARMA with respect to mean squared error of $\hat\beta$ on simulated data for different values of $r$ and $T$.}
	\label{figure:sim:3}
\end{figure}

\begin{figure}[tp!]
	\centering
    \includegraphics[width=1\linewidth]{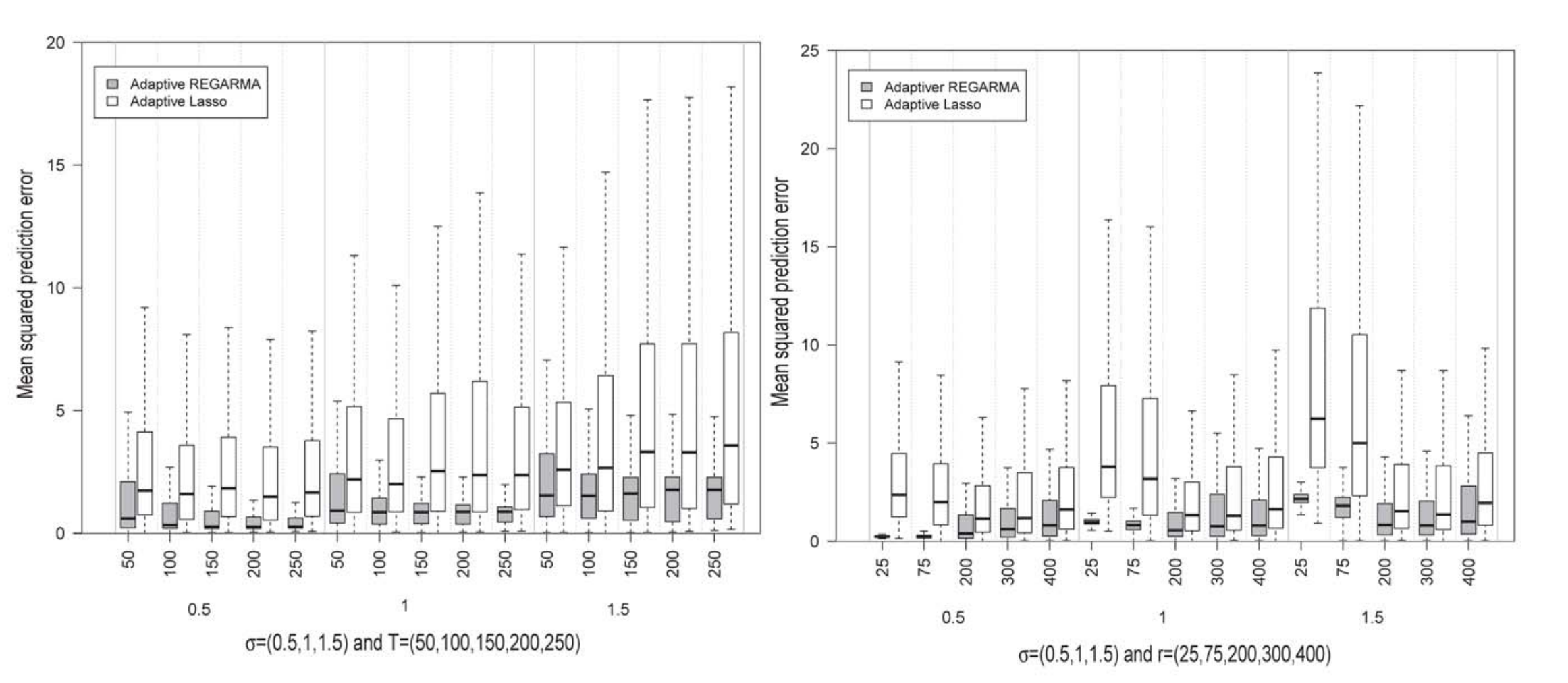}
	\caption{Comparison of adaptive Lasso and adaptive REGARMA with respect to mean squared prediction error on simulated data for different values of $\sigma$, $r$ and $T$.}
	\label{figure:sim:4}
\end{figure}

Figure (\ref{figure:sim:1}) to (\ref{figure:sim:4}) compare adaptive REGARMA and adaptive Lasso with respect to mean squared prediction error, BIC and mean squared error of $\hat \beta$ for $n=50,100,150,200,250$, $\sigma=0.5,1,1.5$ and $r=25,75,200,300,400$. The figures show overall how REGARMA dominates lasso both for low and high-dimensional problems. Figure (\ref{figure:sim:1}) shows that as the number of data points $T$ increases, the relative outperformance of REGARMA versus lasso with respect to MSPE increases. Figures (\ref{figure:sim:2}) shows how REGARMA achieves lower BIC values than lasso,  particularly when $T<r$ but also for some high-dimensional cases. Figure (\ref{figure:sim:3}) compares REGARMA and lasso with respect to the mean squared error of $\hat{\beta}$, averaged over the different regression coefficients. This plot shows the advantage of using REGARMA on time-dependent data in comparison with lasso. Finally, Figure (\ref{figure:sim:4}) shows an outperformance of REGARMA over lasso, regardless of the level of noise $\sigma$.

\section{{Real data analysis}} \label{RealData}
\begin{figure}
	\centering
    \includegraphics[width=.75\linewidth]{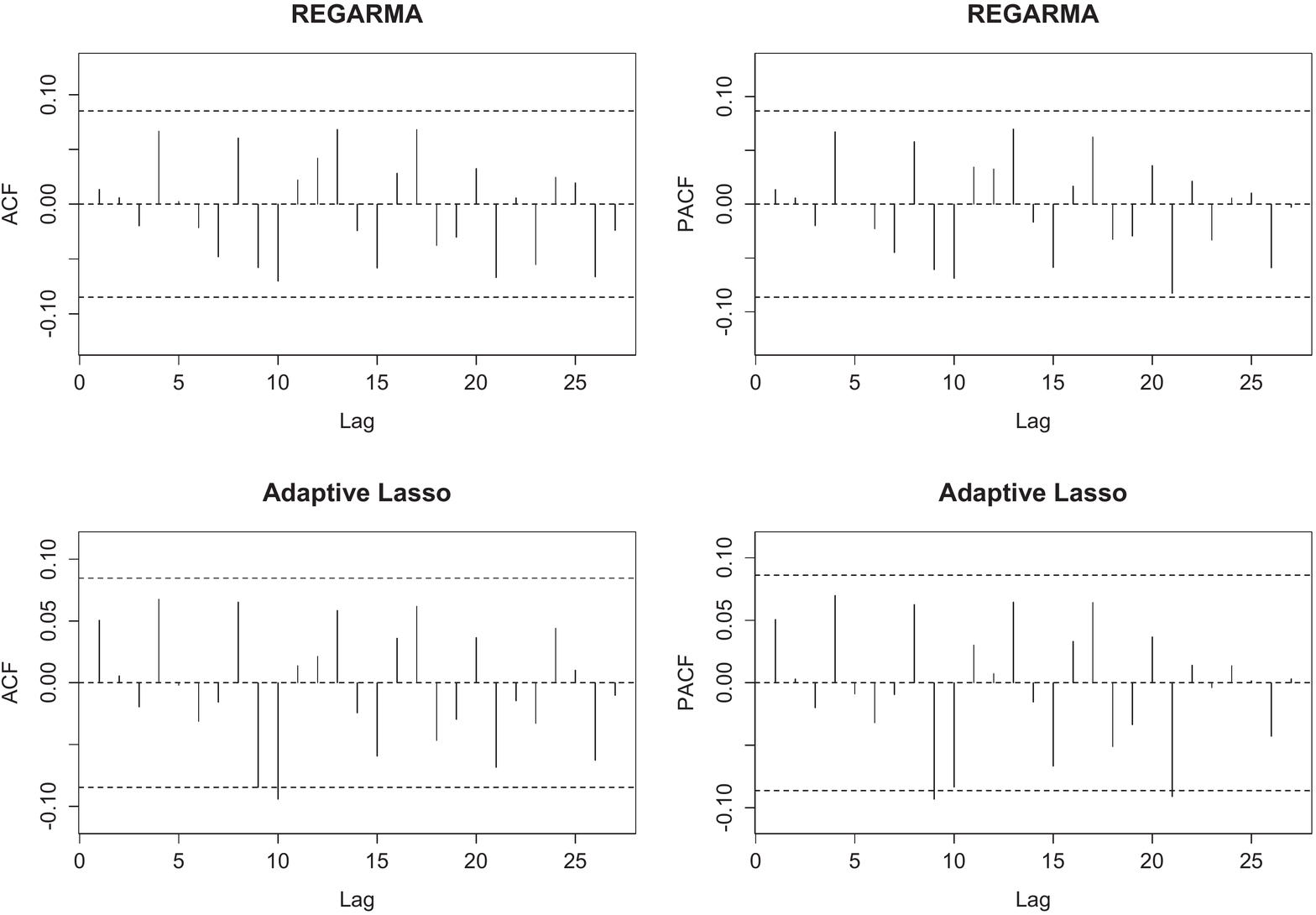}
	\includegraphics[width=.75\linewidth]{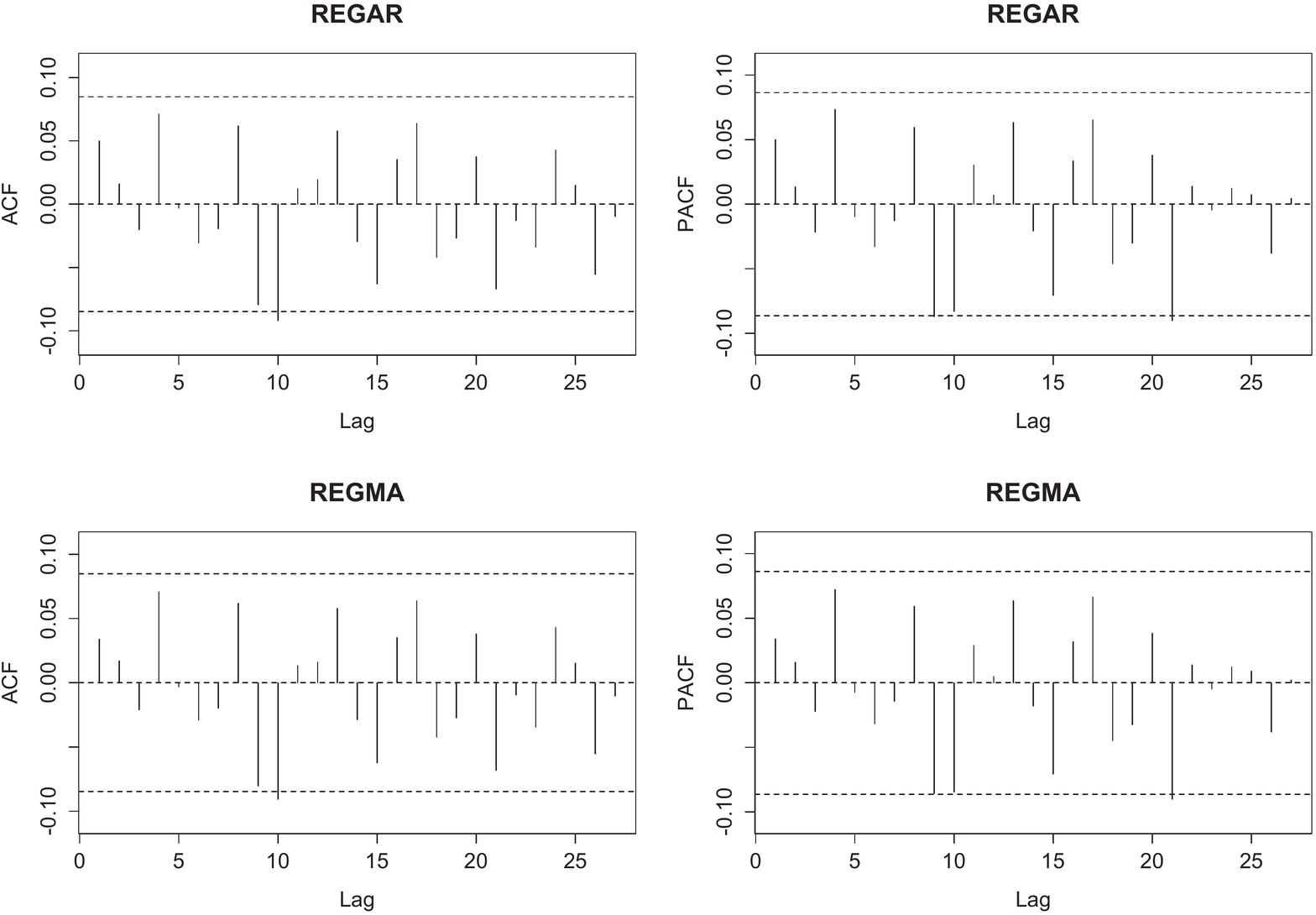}
	\caption{Residual analysis of Adaptive-Lasso, REGAR, REGMA and REGARMA on Istanbul stock exchange data.}
	\label{figure:rf5}
\end{figure}

\normalsize
For the first application, we consider REGARMA in a low-dimensional problem. In particular, we consider financial data on daily returns of the \textit{Istanbul Stock Exchange}(ISE) with seven other international indices, \textit{SP, DAX, FTSE, NIKKEI, BOVESPA, MSCE EU, MSCI EM}, for a period of two years from 2009 to 2011.
The data are publicly available at \url{http://archive.ics.uci.edu/ml} and are considered also by \cite{akbilgic2013novel}. The goal of the analysis is to detect the most effective indices in relation to the ISE index.\\
We set a maximum order of 4 for both $p$ and $q$ and use BIC to select the optimal penalty parameters (i.e. method \textit{b} on page \pageref{method b page number}). Table (\ref{table:rf2}) shows a comparison of REGARMA with adaptive lasso. For REGARMA, we consider also the sub-model with only autoregressive terms, REGAR, and the one with only moving average terms, REGMA (which is essentially the model of \cite{wang2007}), as well as the full REGARMA model.  All four models choose BOVESPA, EU and EM  as the most effective indices for the Istanbul exchange market. These are within the 6 variables selected by \cite{akbilgic2013novel}. From Table (\ref{table:rf2}) and the residual analysis in Figure (\ref{figure:rf5}), we can conclude that the REGARMA family shows a better performance with respect to Mean Squared Error (MSE), Mean Absolute Error (MAE) and BIC compared to adaptive lasso.
\begin{table}
	\caption{\label{table:rf2} Comparison of adaptive lasso and REGARMA models on Istanbul stock exchange (ISE) data.}
	\centering
	\resizebox{1.0\textwidth}{!}{%
	\begin{tabular}{*{5}{l|cccc}}
	MAX AR-MA orders: (4,4)   &  ADAPTIVE-LASSO &  REGAR(2)    	 & REGMA(1) 			& REGARMA(2,1)   \\ \hline
    MEAN SQUARED ERROR  	& 0.4194	& 0.4191	& 0.4192	& 0.4079 \\
    MEAN ABSOLUTE ERROR 	& 0.4932	& 0.4927	& 0.4927	& 0.4907 \\
    BIC       	& 552.44	& 549.12	& 549.12	& 541.41 \\
	\end{tabular}
	}
\end{table}
\newline

For a hight dimensional example, we consider S\&P500 indices. S\&P500 is one of the leading stock market index for US equity: it is based on 500 leading companies and captures approximately 80\% coverage of the available market capitalization. The goal of the analysis is to find the S\&P500 indices most related to the \textit{AT\&T Inc} index based on monthly data in a period of fourteen years from 2000 to 2014. These data are publicly available at \url{http://thomsonreuters.com/}.
After removing variables with the majority of missing values, the dataset contains 416 variables and 170 datapoints. As before, we apply adaptive lasso and adaptive REGARMA to these data and choose the optimal penalties by BIC. Moreover, we set a maximum order of 4 for both $p$ and $q$ and let the model choose the optimal orders (using method \textit{b} on page \pageref{method b page number}).\\

\begin{figure}
	\centering
	\includegraphics[scale=.40]{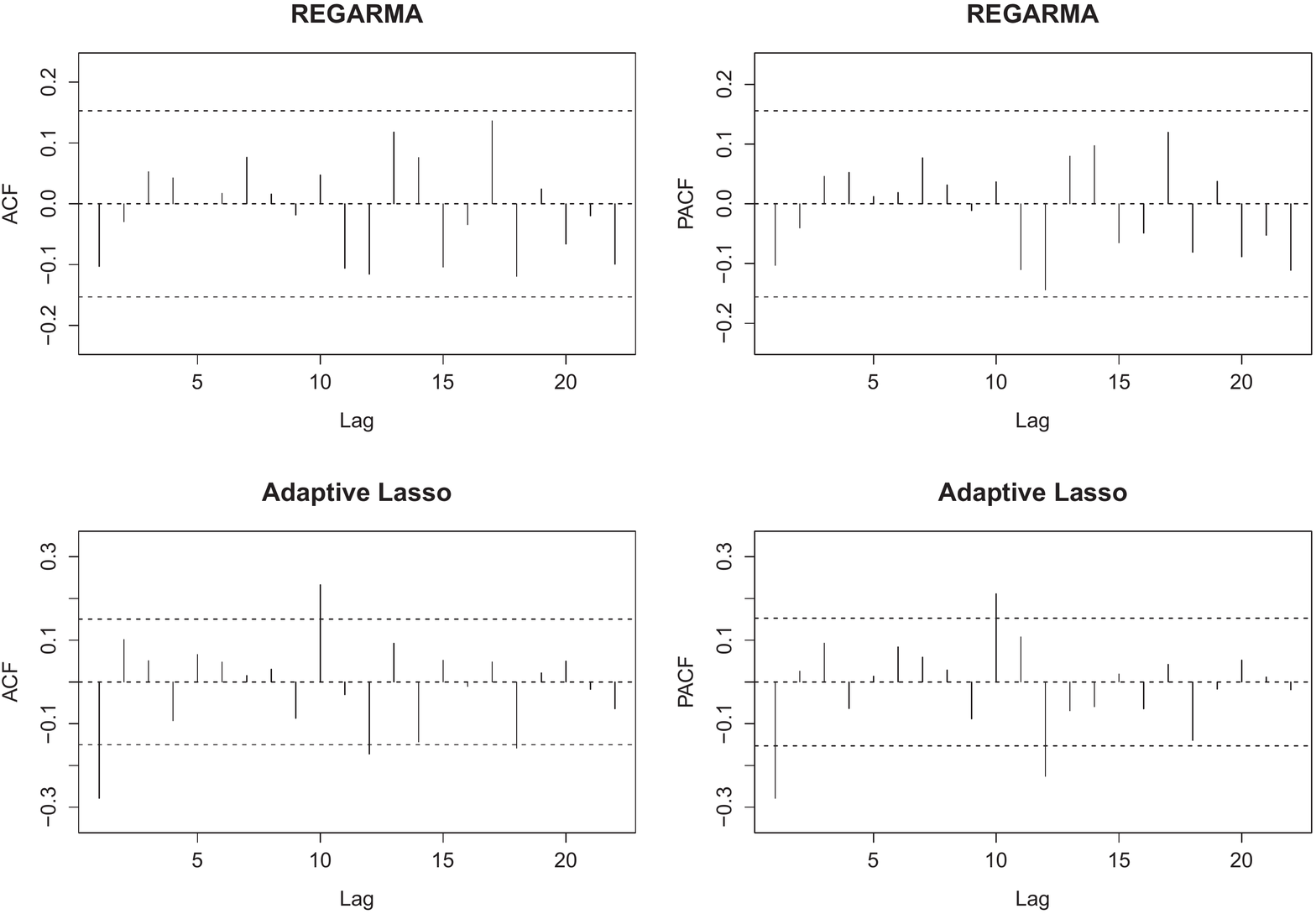}
	\includegraphics[scale=.40]{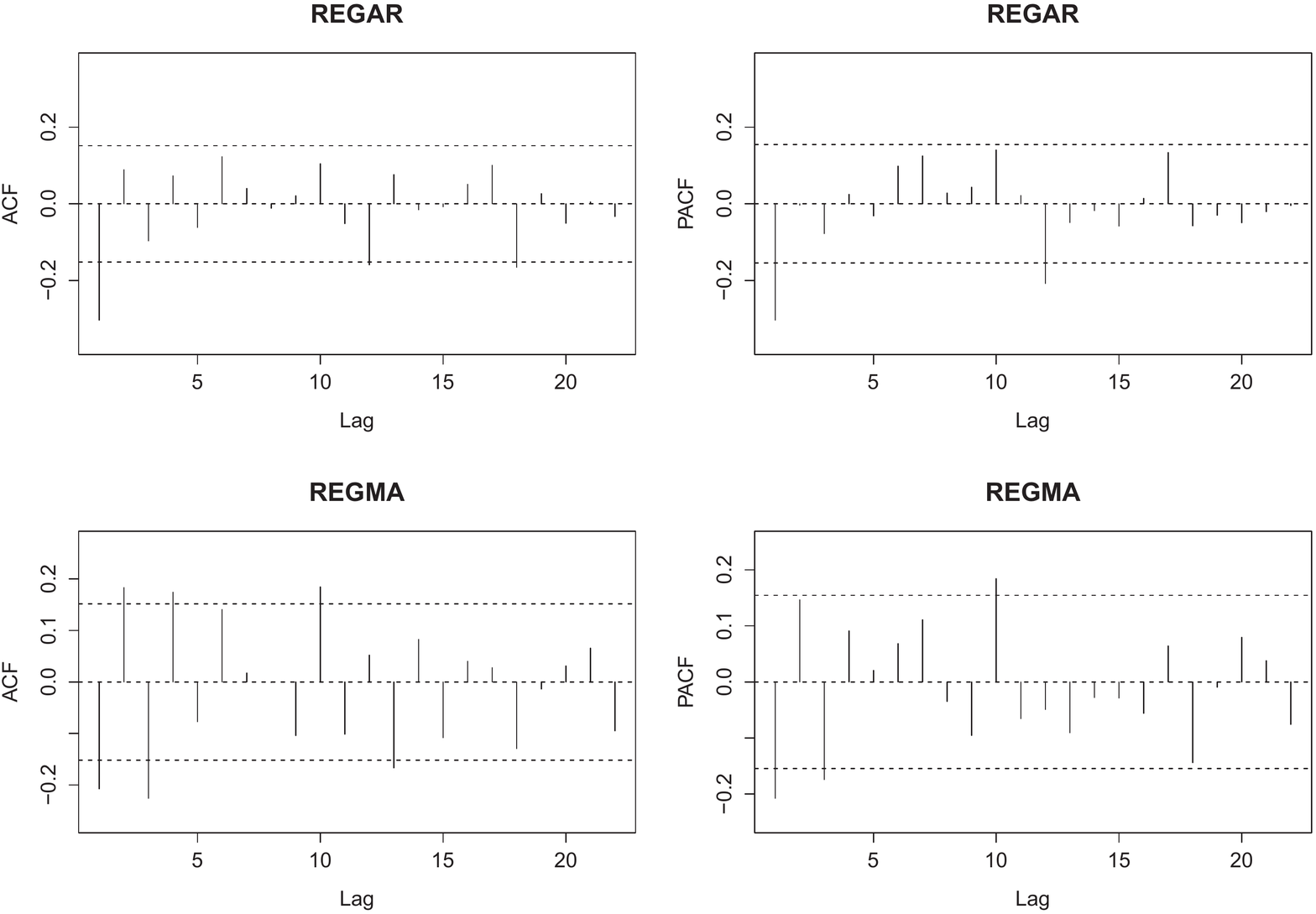}
	\caption{Residual analysis of Adaptive-Lasso, REGAR, REGMA and REGARMA on AT\&T index.}
	\label{figure:rf6}
\end{figure}

Table (\ref{table:rf3}) summarises the results in terms of MSE, MAE, BIC and the number of non-zero coefficients. Moreover, Figure (\ref{figure:rf6}) illustrates the residual analysis of these four models. Both Table (\ref{table:rf3}) and Figure (\ref{figure:rf6}) show an improved performance of REGARMA compared to the other methods.
\normalsize
\begin{table}
	\caption{\label{table:rf3} Comparison of adaptive lasso and  REGARMA models  on \textit{AT\&T Inc} and S\&P500 data.}
	\centering
	\resizebox{1\textwidth}{!}{%
		\begin{tabular}{*{5}{l|cccc}}
				MAX AR-MA orders: (4, 4)   &  ADAPTIVE-LASSO &  REGAR(2)    	 & REGMA(1) 			& REGARMA(2,3)   \\ \hline
				MEAN SQUARED ERROR  	& 1.34	& 1.84	& 4.18	& .083 \\
				MEAN ABSOLUTE ERROR	& 87.01	& 112.30	& 182.07	& 75.34 \\
						BIC      	& 28.83	& 27.81	& 31.2	& 22.63 \\
				NON-ZERO COEFFICIENTS &    272 & 266 & 282 & 263
		\end{tabular}
	}
\end{table}	
\section{{Conclusion}}\label{Conclusion and discussion}

In this paper we extend the idea of regression-time series models proposed in~\cite{wang2007} to a more general class of models, thus covering a wide spectrum of applications involving multivariate time-dependent data.
In particular, we study an autoregressive moving average model with time-dependent explanatory variables and present $l_1$ penalised inference for the estimation of its parameters. Our model lends itself naturally to parameter estimation and implementation, contrary to  the linear regression with ARMA errors of \cite{wu2012shrinkage}.
 We prove asymptotic properties of the proposed model in low and high dimensional situations, with the latter not considered by the existing literature on time-series regression models.
We test the performance of the model on a simulation study and show a successful application on financial data.

\normalsize
\section{{Appendix (Proof of theorems)}}

\begin{t6}[Theorem \ref{theorem2}]
	Assume $\lambda_n\sqrt{n} \rightarrow \lambda_\circ, \text{  }
	\gamma_n\sqrt{n} \rightarrow \gamma_\circ,  \text{  }
	\tau_n\sqrt{n} \rightarrow \tau_\circ
	$, and $\delta=(u',v',w')$. Define
	\begin{equation}\label{theorem2:eq1}
		k_n(\delta)=Q_n(\Theta^\circ +n^{-(1/2)} \delta)-Q_n(\Theta^\circ ).
	\end{equation}
	Note that $k_n$ achieves a minimum at $\delta=\sqrt{n}(\hat\Theta-\Theta^\circ)$. Using (\ref{openmodel1}), it implies that
	\begin{subequations}
	\begin{align}
		k_n(\delta)=
		             & \bigg(L_n(\Theta^\circ +\frac{\delta}{\sqrt{n}})-L_n(\Theta^\circ )\bigg)          \label{t2eq:eq1}  \\
		             & +(n\lambda'_n|\beta^\circ +\frac{u}{\sqrt{n}}|-n\lambda'_n|\beta^\circ |) \label{t2eq:eq2} \\
		             & +(n\gamma'_n|\phi^\circ +\frac{v}{\sqrt{n}}|-n\gamma'_n|\phi^\circ |) \label{t2eq:eq3}     \\
		             & +(n\tau'_n|\theta^\circ +\frac{w}{\sqrt{n}}|-n\tau'_n|\theta^\circ |).
		\label{t2eq:eq4}
	\end{align}
	\end{subequations}
	The last three terms can be simplified as
\begin{align*}
			(n\lambda'_n|\beta^\circ +\frac{u}{\sqrt{n}}|-n\lambda'_n|\beta^\circ |) & =
			\left(
			\sqrt{n}u\lambda'_n(
			\frac{|\beta^\circ +u/\sqrt{n}|-|\beta^\circ |}{u/\sqrt{n}})
			\right)\\
			 & \underset{n \rightarrow\infty}{\rightarrow}
			\lambda_\circ\sum_{i=1}^{r}\{(u_i sign(\beta^\circ _i)I(\beta^\circ _i\neq 0))+|u_i|I(\beta_i^\circ =0)  \}.
\end{align*}
Similarly, for the other two terms, we have
\begin{align*}
			(\ref{t2eq:eq3})                                               & \underset{n \rightarrow\infty}{\rightarrow} \gamma_\circ\sum_{j=1}^{p}\{(v_j sign(\phi^\circ _j)I(\phi^\circ _j\neq 0))+|v_j|I(\phi_j^\circ =0)  \}      \\
			(\ref{t2eq:eq4})                                               & \underset{n \rightarrow\infty}{\rightarrow} \tau_\circ\sum_{k=1}^{q}\{(w_k sign(\theta^\circ _k)I(\theta^\circ _k\neq 0))+|w_k|I(\theta_k^\circ =0)  \}.
\end{align*}
For (\ref{t2eq:eq1}), we have
	\begin{equation*}
		\begin{aligned}
			(\ref{t2eq:eq1})= & -e'e  +\left((Y-H_{(q)}  \theta^\circ -H_{(p)}\phi^\circ -X'\beta^\circ )-(X',H_{(p)},H_{(q)})\frac{\delta}{\sqrt{{n}}})\right)' \\
			\times
			 & \left((Y-H_{(q)} \theta^\circ -H_{(p)}\phi^\circ -X'\beta^\circ )-(X',H_{(p)},H_{(q)})\frac{\delta}{\sqrt{{n}}})\right).
		\end{aligned}
	\end{equation*}
Set $A=(X',H_{(p)},H_{(q)})$ and recall that $e=Y-H_{(q)}\theta^\circ -H_{(p)}\phi^\circ -X'\beta^\circ $. Then, we have
	\begin{equation*}
		\begin{aligned}
			Q_n(\Theta^\circ  & +\frac{\delta  }{\sqrt{n}})-Q_n(\Theta^\circ )
			              =(e'-\frac{\delta'}{\sqrt{n}}A')
			(e-A\frac{\delta}{\sqrt{n}})
			-e'
			e+(\ref{t2eq:eq2})+(\ref{t2eq:eq3})+(\ref{t2eq:eq4}).
		\end{aligned}
	\end{equation*}
	The right-hand side of the last equation is equivalent to
	\begin{equation}
		(\frac{\delta'A'}{\sqrt{n}})
		(\frac{A\delta}{\sqrt{n}})-
		(\frac{\delta'A'}{\sqrt{n}})
		e-
		e'
		(\frac{A\delta}{\sqrt{n}})
		+(\ref{t2eq:eq2})+(\ref{t2eq:eq3})+(\ref{t2eq:eq4})
		\label{t2eq:eq5}.
	\end{equation}
	From left to right, we now prove that the first term in (\ref{t2eq:eq5}) is bounded and the two other terms follow (asymptotically) normal distributions, i.e.

	\begin{align}
		                                                 & (\frac{\delta'A'}{\sqrt{n}})
		(\frac{A\delta}{\sqrt{n}}) \rightarrow O(1)      & \label{theorem2:eq6}         \\
		                                                 & (\frac{\delta'A'}{\sqrt{n}})
		e \rightarrow f_1                                & \label{theorem2:eq7}         \\
		                                                 & e'
		(\frac{A\delta}{\sqrt{n}}) \rightarrow f_1'=f_1. & \label{theorem2:eq8}
	\end{align}
		
Let $H'_2=\frac{A'}{\sqrt{n}}$. Then,
\begin{align*}
 &  H'_2  e=\frac{1}{\sqrt{n}}(X',H_{(p)},H_{(q)})'e\\
 &\sqrt{n}H'_2e=(X',H_{(p)},H_{(q)})'e\\
 &H^\circ_{t}=\sqrt{n}H'_{2_t} e_t=(X'_t,H_{(p)_t},H_{(q)_t})'e_t.
\end{align*}
$H_t^\circ$ is a martingale difference sequence(in short \textit{mds}) because
\begin{align*}
\mathbb{E}(H^\circ_t \vert t=t-1,t-2,\ldots,t-(p+q))& =\mathbb{E}((X'_t,H_{(p)_t},H_{(q)_t})'e_t|<t) \\
&=(X'_t,H_{(p)_t},H_{(q)_t})'\mathbb{E}(e_t)=0.
\end{align*}
In order to establish that the conditions of the mds central limit theorem are satisfied (refer to ~\citet[p 51]{martin2012econometric} for the mds central limit theorem and conditions), define
\begin{align*}
& \bar{\mu}=\frac{1}{n}\sum_{t=T_0+1}^{T}H^\circ_t, \quad
 \bar{\sigma}^2=\frac{1}{n}\sum_{t=T_0+1}^{T}Var(H^\circ_t)=\sigma^4 U_B.
\end{align*}
To show the boundedness condition in the martingales central limit theorem, choose $\delta=2$, so that
\begin{align*}
\mathbb{E}(|H^\circ_t|^4)=
\mathbb{E}(e_t^4)
E \left(X'_t,H_{(p)_t},H_{(q)_t}\right)^4.
\end{align*}
Under assumption $a$, $\mathbb{E}(e_t^4)<\infty$ and it can be shown that $E \left(X'_t,H_{(p)_t},H_{(q)_t}\right)^4<\infty$, provided $y_t$ and $x_t$ are stationary and ergodic. Moreover
\small
\begin{align}\label{Martingle CLT-eq:1}
\frac{1}{n}& \sum_{t=T_0+1}^{T} e_t^2 \left(X_t,H'_{(p)_t},  H'_{(q)_t}\right)^2= \nonumber \\
&\quad \frac{1}{n}\sum_{t=T_0+1}^{T}(e_t^2-\sigma^2) \left(X'_t,H_{(p)_t},H_{(q)_t}\right)^2+
\sigma^2\frac{1}{n}\sum_{t=T_0+1}^{T} \left(X'_t,H_{(p)_t},H_{(q)_t}\right)^2.
\end{align}
\normalsize
The first term in (\ref{Martingle CLT-eq:1}) is a mds, which has mean zero. So using the weak law of large numbers, we have that $\frac{1}{n}\sum_{t=T_0+1}^{T}(e_t^2-\sigma^2) \left(X'_t,H_{(p)_t},H_{(q)_t}\right)^2 \overset{p}{\rightarrow}0$. The second term in the right hand side of (\ref{Martingle CLT-eq:1}) also tends to $\sigma^2U_B$. As a result
\begin{align*}
\frac{1}{n}\sum_{t=T_0+1}^{T}e_t^2 &\left(X_t,H'_{(p)_t},  H'_{(q)_t}\right)^2\overset{p}{\rightarrow}\sigma^2U_B.
\end{align*}
Therefore, by the central limit theorem for martingales, it follows that $\frac{1}{\sqrt{n}}H'_2e \overset{d}{\rightarrow} N(0,\sigma^2 U_B)$
and
\begin{equation*}
	(\frac{\delta'A'}{\sqrt{n}})e                               \overset{d}{\rightarrow} \delta' W,
\end{equation*}
where $\delta=(u',v',w')$ and $W\sim MVN(O,\sigma^{2} U_B)$. Then
\begin{equation*}
	-\left( (\ref{theorem2:eq7})+(\ref{theorem2:eq8}) \right) \overset{d}{\rightarrow}
	 -2\delta' W.
\end{equation*}
If $X_i,i=1,2,3,\ldots,r$ and $y_t$ are stationary and ergodic, it is possible to show that (\ref{theorem2:eq6}) tends to $\delta'U_B\delta$
where $U_B$ is the covariance matrix of $(X',H_{(p)},H_{(q)})$, i.e. (\ref{theorem2:eq6})$\rightarrow O(1)$.
Finally,  $k_n(\delta)$ in equation (\ref{theorem2:eq1}) converges to
\begin{equation*}
	\begin{aligned}
		k_n(\delta) \overset{d}{\rightarrow} -2\delta' N(O,\sigma^{2}U_B)+\delta'U_B\delta & +
		\lambda_\circ\sum_{i=1}^{r}\{(u_i sgn(\beta^\circ _i)I(\beta^\circ _i\neq 0))+|u_i|I(\beta_i^\circ =0)  \} \\
		& +\gamma_\circ\sum_{j=1}^{p}\{(v_j sgn(\phi^\circ _j)I(\phi^\circ _j\neq 0))+|v_i|I(\phi_j^\circ =0)  \}       \\
		& +\tau_\circ\sum_{k=1}^{q}\{(w_k sgn(\theta^\circ _k)I(\theta^\circ _k\neq 0))+|w_k|I(\theta_k^\circ =0)  \} .
	\end{aligned}
\end{equation*}
 Up to here, we have proved that $k_n(\delta)\overset{d}{\rightarrow}k(\delta)$. To show that $\arg\min(k_n(\delta))=\sqrt{n}(\hat{\Theta}-\Theta^\circ)\overset{d}{\rightarrow}\arg\min(k(\delta))$ it is enough to prove that $\arg\min\{k_n(\delta)\}= O_p(1)$~\citep{kim1990,knight2000}. To show this, note that
 \small
\begin{equation*}
	\begin{aligned}
		k_n(\delta) & =(\frac{\delta'A'}{\sqrt{n}})
		(\frac{A\delta}{\sqrt{n}})-
		(\frac{\delta'A'}{\sqrt{n}})
		e-
		e'
		(\frac{A\delta}{\sqrt{n}})+ \\
		& (n\lambda'_n|\beta^\circ+\frac{u}{\sqrt{n}}|-n\lambda'_n|\beta^\circ|) +
		(n\gamma'_n|\phi^\circ+\frac{v}{\sqrt{n}}|-n\gamma'_n|\phi^\circ|) +
		(n\tau'_n|\theta^\circ+\frac{w}{\sqrt{n}}|-n\tau'_n|\theta^\circ|)  \\
		& \geq
		(\frac{\delta'A'}{\sqrt{n}})
		(\frac{A\delta}{\sqrt{n}})-
		(\frac{\delta'A'}{\sqrt{n}})
		e-
		e'
		(\frac{A\delta}{\sqrt{n}})-
		(n\lambda'_n|un^{-1/2}| -
		(n\gamma'_n|vn^{-1/2}|) -
		(n\tau'_n|wn^{-1/2}|)  \\
		& \geq
		(\frac{\delta'A'}{\sqrt{n}})
		(\frac{A\delta}{\sqrt{n}})-
		(\frac{\delta'A'}{\sqrt{n}})
		e-
		e'
		(\frac{A\delta}{\sqrt{n}})
		-
		(\lambda'_\circ+\epsilon_0)|u|-(\gamma'_\circ+\epsilon_0)|v|-(\tau'_\circ+\epsilon_0)|w|+f_n(\delta) \\
		& =k^*_n(\delta)
	\end{aligned}
\end{equation*}
\normalsize
where $\epsilon_0>0$ is a vector of positive constants. The fourth term in $k^*_n(\delta)$ for example, comes from the fact that $\forall\epsilon_0>0$, there exists an $N$ such that  $\text{\textit{if }}\, n\geq N$, then $|\lambda^\circ-\sqrt{n}\lambda_n|<\epsilon_0$. Then, $\sqrt{n}\lambda_n<\lambda^\circ+\epsilon_0$. In addition, $k_n(0)=k^*_n(0)$ and $f_n(\delta)=o_p(1)$.
As a result $\arg\min\{k^*_n(\delta)\}=O_p(1)$ and $\arg\min\{k_n(\delta)\}=O_p(1)$. The proof of the theorem is completed.
\end{t6}

\begin{t6}[Theorem \ref{theorem3}]
	Let $\alpha_n=n^{-1/2}+a_n$, and $\{\Theta^\circ +\alpha_n\delta:||\delta||\leq d,\delta=(u,v,w)'\}$ be a ball around $\Theta^\circ $. Then for $||\delta||=d$ we have
	\small
	\begin{equation*}
		\begin{aligned}
			R_n(\delta)&=Q^*_n(\Theta^\circ +\alpha_n\delta)-Q^*(\Theta^\circ ) \\
			& \geq L_n(\Theta^\circ +\alpha_n\delta)-L_n(\Theta^\circ )+K \\
			& \geq L_n(\Theta^\circ +\alpha_n\delta)-L_n(\Theta^\circ )+K' \\
			& \geq L_n(\Theta^\circ +\alpha_n\delta)-L_n(\Theta^\circ )+K'' \\
			\text{where} \\
			K=& n\sum_{i \in s_1}^{}\lambda^*_i(|\beta^\circ _i+\alpha_nu_i|-|\beta^\circ _i|)+n\sum_{j \in s_2}^{}\gamma^*_j(|\phi_j^\circ +\alpha_nv_j|-|\phi_j^\circ |) +n\sum_{k \in s_3}^{}\tau^*_k(|\theta_k^\circ +\alpha_nw_k|-|\theta_k^\circ |) \\
		\end{aligned}
	\end{equation*}
	\begin{align}
			\text{(Using triangular inequality)}:K'   &= -n\alpha_n \sum_{i \in s_1}^{} \lambda^*_i|u_i|-n\alpha_n \sum_{j \in s_2}^{} \gamma^*_j|v_j|-n\alpha_n \sum_{k \in s_3}^{} \tau^*_k|w_k|  \nonumber\\
			\text{(Penalties $\leq \alpha_n$ by definition)}:K'' & = -n\alpha_n^2(r_\circ+p_\circ+q_\circ)d.      \label{Adaptive-regarma-theorem1:1}
	\end{align}	
		\normalsize
	The last equation holds because of the decreasing speed of $\alpha_n$. Similar calculations to those in theorem (\ref{theorem2}) result in
	\begin{equation}
		L_n(\Theta^\circ+\alpha_n\delta)-L_n(\Theta^\circ) \rightarrow n\alpha_n^2\{\delta' U_B \delta+o_p(1)\}.
		\label{Adaptive-regarma-theorem1:2}
	\end{equation}
	Because (\ref{Adaptive-regarma-theorem1:2}) dominates (\ref{Adaptive-regarma-theorem1:1}), then for any given $\eta>0$ , there is a large constant $d$ such that
	\begin{equation*}
		Pr[\underset{||\delta||=d}{inf}\{Q^*_n(\Theta^\circ +\alpha_n\delta)\}>Q^*_n(\Theta^\circ )] \geq 1-\eta.
	\end{equation*}
	This result shows that with probability at least $1-\eta$, there is a local minimiser in the ball $\{\Theta^\circ +\alpha_n\delta:||\delta|| \leq d\}$ and as a result a minimiser, $Q^*_n(\Theta)$, such that $||\hat \Theta^*-\Theta^\circ||=O_p(\alpha_n)$.
	The proof is completed.
\end{t6}
\begin{t6}[Theorem \ref{theorem4}]
	This proof follows from the fact that $Q_n^*(\hat \Theta^*)$ must satisfy
	\begin{align}
			\frac{\partial Q^*_n(\Theta)}{\partial\beta_i}
			\bigg\vert_{\hat\Theta^*}  & =\frac{\partial L_n(\hat \Theta^*)}{\partial \beta_i}-n\lambda^*_i sign(\hat{\beta}^*_i)  \nonumber \\
			                  &=\frac{\partial L_n( \Theta^\circ )}{\partial\beta_i}+nU_{{i}} (\hat\Theta^*-\Theta^\circ )\{1+o_p(1)\}-n\lambda^*_i sign(\hat\beta^*_i) \label{t4:eq3}
	\end{align}
	where $U_{{i}}$ is th $i^{th}$ row of $U_B$ and $i \in s_1^c$. The second term in (\ref{t4:eq3}) is a direct result of (\ref{adaptive-REGARMA in matrix}) by adding a $\pm X'\beta^\circ,\pm H_{(p)}\phi^\circ$ and $\pm H_{(q)}\theta^\circ$ to  $L_n(\hat \Theta^*)$. By using the central limit theorem, the first term of equation (\ref{t4:eq3}), $\sum_{t}e_tx'_{ti}$, will be of order $O_p(n^{1/2})$ and the second term of order $O_p(n^{1/2})$. Furthermore both are dominated by $n\lambda^*_i$ since $b_n \sqrt n \rightarrow \infty$. Then the sign of $\frac{\partial Q^*_n(\hat\Theta^*)}{\partial\beta_i}$ is dominated by the sign of $\hat \beta^*_i$ and $\hat \beta^*_i=0$ in probability. Analogously, we can show that $Pr(\hat \phi^*_{s_2^c}) \overset{p}{\rightarrow} 1$ and $Pr(\hat \theta^*_{s_3^c}) \overset{p}{\rightarrow} 1$.\\
	The proof is completed.
\end{t6}
\begin{t6}[Theorem \ref{theorem5}]
	From Theorem (\ref{theorem3}) and (\ref{theorem4}) one can conclude that $Pr(\hat \Theta_2^*=0) \overset{p}{\rightarrow}1$. Thus, $Q_n^*(\Theta) \overset{with \quad pr \rightarrow 1}{\xrightarrow{\hspace*{1.5cm}}}Q^*_n(\Theta_1)$. So it implies that the Lasso estimator $\hat\Theta^*_1$ satisfies the equation
	\begin{equation*}
		\frac{\partial Q_n^* (\Theta_1)}{\partial \Theta_1}|_{\Theta_1=\hat\Theta^*_1}=0.
	\end{equation*}
	From Theorem (\ref{theorem3}), $\hat \Theta_1^*$ is a $\sqrt{n}-consistent$ estimator, thus a Taylor expansion of the above equation yields
	\begin{equation*}
		\begin{aligned}
			0 & =\frac{1}{\sqrt{n}}\frac{\partial L_n (\hat \Theta_1^*)}{\partial \Theta_1}+F(\hat\Theta_1^*)\sqrt n                                                  \\
			  & =\frac{1}{\sqrt{n}}\frac{\partial L_n (\hat \Theta_1^\circ )}{\partial \Theta_1}+F(\hat\Theta_1^\circ )\sqrt n + U_0 \sqrt{n}(\hat\Theta_1^*-\Theta_1^\circ )+o_p(1),
		\end{aligned}
	\end{equation*}
	where F is the first-order derivation of the tuning function
	\begin{equation*}
		\sum_{i \in s_1}\lambda_i|\beta_i|+
		\sum_{j \in s_2}\gamma_j|\phi_j|+
		\sum_{k \in s_3}\tau_k|\theta_k|.
	\end{equation*}
	For $n$ sufficiently large, $F(\hat\Theta_1^*)=F(\Theta_1^\circ )$, thus
	\begin{equation*}
		\begin{aligned}
			(\Theta_1^\circ -\hat\Theta_1^*)\sqrt{n} & =\frac{U_0^{-1}}{\sqrt{n}}\frac{\partial L_n(\Theta_1^\circ )}{\partial \Theta_1}+o_p(1) \\
			                                    & \overset{d}{\rightarrow} N(0,\sigma^2 U_0^{-1}).
		\end{aligned}
	\end{equation*}
	The proof is completed.
\end{t6}
\begin{t6}[Theorem \ref{theorem:Consistency}]
Let $Y=(y_1,y_2,\ldots,y_n)'$, $ \hat{Y} =(\hat y_1^{}, \hat y_2^{},\ldots, \hat y_n^{})'$ and $Y^\circ=(y_1^\circ,y_2^\circ,\ldots,y_n^\circ)'$ be observations, REGARMA predictions and predictions from the true model, respectively, that is
\begin{align*}
&  \hat{Y}  =X'\hat\beta^{}+H_{(p)}\hat\phi^{}+H_{(q)}\hat\theta^{}\\
& Y^\circ =X'\beta^\circ+H_{(p)}\phi^\circ+H_{(q)}\theta^\circ.
\end{align*}
Define a set $C=\{
X'\beta+H_{(p)}\phi+H_{(q)}\theta
; $ $
\sum_{j=1}^{r}|\beta_j|\leq  K_\lambda,
\sum_{k=1}^{p}|\phi_k|\leq K_\gamma,
\sum_{l=1}^{q}|\theta_l|\leq K_\tau
\}$.
Note that $C$ is a compact and convex subset of $\mathbb{R}^n$ and that $ \hat{Y} $ is the projection of $Y$ on $C$. As a result of the convexity of $C$, for any vector $v$ in $C$, the vector $v- \hat{Y} $ must be at an obtuse angle to the vector $Y- \hat{Y} $. This means that
\begin{equation*}\label{consistency:eq2}
\bigg\langle(v- \hat{Y} ),(Y- \hat{Y} )\bigg\rangle \leq 0.
\end{equation*}
Since $Y^\circ\in C$, then the inner product of $(Y^\circ- \hat{Y} )$ and $(Y- \hat{Y} )$ is non-positive
\begin{equation}\label{consistency:eq3}
\bigg\langle(Y^\circ- \hat{Y} ),(Y- \hat{Y} )\bigg\rangle\leq 0.
\end{equation}
Using (\ref{consistency:eq3}) and some simple algebra we have
\fontsize{10pt}{10pt}
\begin{align}
\widehat{ MSPE}&=\frac{1}{n}  \Vert Y^\circ- \hat{Y}   \Vert^2  \nonumber \\
& \leq \frac{1}{n}\bigg\langle(Y-Y^\circ),( \hat{Y} -Y^\circ)\bigg\rangle \nonumber\\
& \leq
  \frac{1}{n}e'
  \left(
   X'(\hat\beta^{}-\beta^\circ) +
   H_{(p)}(\hat\phi^{}-\phi^\circ) +
   H_{(q)}(\hat\theta^{}-\theta^\circ)
  \right) \nonumber\\
& =
  \frac{1}{n}\left(e' X'(\hat\beta^{}-\beta^\circ) +
  e' H_{(p)}(\hat\phi^{}-\phi^\circ) +
  e' H_{(q)}(\hat\theta^{}-\theta^\circ)
  \right) \nonumber\\
& \leq
\frac{1}{n}\left(
  2K_\lambda \max_{1 \leq i \leq r}|e' X_i|+
    2K_\gamma \max_{1 \leq j \leq p}|e' H_{{(p)}_j}|+
      2K_\tau \max_{1 \leq k \leq q}|e' H_{{(q)}_k}|
      \right)
      .
       \label{consistency:eq4}
\end{align}
\normalsize
On the other hand, conditioning on $X$ and history of $y$ results in
\begin{align*}
& e'X'\sim N(O,\sigma^2 (XX'))\\
& e'H_{(p)}\sim N(O,\sigma^2 H'_{(p)}H_{(p)})\\
& e'H_{(q)}\sim N(O,\sigma^2 (H'_{(q)}H_{(q)})).
\end{align*}
Using Lemma 3 in ~\cite{sourvan2013},
\begin{align}
& \mathbb{E}(\underset{i=1,2,3,\ldots,r}{ \max|e'{X_i}|)} \leq M_1\sigma\sqrt{2n\log(2r)}  		\label{consistency:eq5} \\
& \mathbb{E}(\underset{j=1,2,3,\ldots,p}{\max|e'H_{(p)_j}|)} \leq M_2\sigma\sqrt{2n\log(2p)}   \label{consistency:eq6} \\
& \mathbb{E}(\underset{k=1,2,3,\ldots,q}{\max|e'H_{(q)_k}|)}  \leq M_3\sigma\sqrt{2n\log(2q)}  \label{consistency:eq7}
.
\end{align}
Substituting (\ref{consistency:eq5},\ref{consistency:eq6},\ref{consistency:eq7}) in (\ref{consistency:eq4}) result in
\small
\begin{align*}
\frac{1}{n}\Vert Y^\circ- \hat{Y}  \Vert^2
\leq
\frac{1}{n}\left(
 2K_\lambda M_1\sigma\sqrt{2n\log(2r)}+
 2K_\gamma M_2\sigma\sqrt{2n\log(2p)}+
 2K_\tau M_3\sigma\sqrt{2n\log(2q)}
 \right)
 .
\end{align*}
\normalsize
But $M_{max}=\sup \{M_1,M_2,M_3 \}$ and $K_{max}=\sup\{K_\lambda,K_\gamma,K_\tau\}$, therefore
\begin{equation*}\label{consistency:eq8}
\frac{1}{n}\Vert Y^\circ- \hat{Y}  \Vert^2
\leq
 \frac{2K_{max} M_{max}\sigma}{n}\bigg(\sqrt{2n\log(2r)}+
\sqrt{2n\log(2p)}+
\sqrt{2n\log(2q)}\bigg)
\end{equation*}
and
\fontsize{10pt}{10pt}
\begin{equation}\label{consistency:eq9}
\widehat{ MSPE}(\hat\beta^{},\hat\phi^{},\hat\theta^{})
 \leq
 \frac{ 2K_{max} M\sigma}{\sqrt{n}}
 \left(\sqrt{2\log(2r)}+
\sqrt{2\log(2p)}+
\sqrt{2\log(2q)}
\right).
\end{equation}
\normalsize
The proof of the theorem is completed.
\end{t6}
\begin{t4}\label{lemma2}
If $X_1,X_2,X_3,\ldots,X_m$ are $m$ {dependent} mean zero random variables where $|X_i|\leq L,\forall i\in\{1,2,3,\ldots,m\}$. Then, $\forall \beta \in \mathbb{R}$,
\begin{align*}
\mathbb{E}(e^{\beta\sum_{i=1}^{m}x_i}) \leq  e^{(mL\beta)^2}.
\end{align*}
\end{t4}
\normalsize
\begin{proof}
This result extends the result of \cite{sourvan2013} from independent variables to dependent variables.

It is obvious that $\sum_{i=1}^{m}x_i\leq mL$ then,
$$
\mathbb{E}(e^{\beta\sum_{i=1}^{m}x_i})=\int_{-mL}^{mL}e^{\beta\sum_{i=1}^{m}x_i} d\mu(\sum_{i=1}^{m}x_i),
$$
where $\mu$ is a probability distribution. On the other hand, $x\mapsto e^{x\beta}$ is a convex map. Define $t=\frac{\sum_{i=1}^{m}{x_i}}{2mL}+\frac{1}{2}$, then 
$$
e^{\beta\sum_{i=1}^{m}x_i}= e^{\beta\bigg(t( mL) -(1-t)(mL)\bigg)}\leq te^{\beta mL}+(1-t)e^{-\beta m L}.
$$
Using the fact that $\mathbb{E}(\sum_{i=1}^{m} x_i)=\sum_{i=1}^{m}\mathbb{E}(x_i)=0$,
\begin{align*}
\int_{-mL}^{mL} e^{\beta\bigg(t( mL) -(1-t)(mL)\bigg)}d\mu(\sum_{i=1}^{m}x_i)
\leq &
\int_{-mL}^{mL} te^{\beta mL}+(1-t)e^{-\beta m L}d\mu(\sum_{i=1}^{m}x_i)\\
& =\frac{e^{\beta mL}+e^{-\beta mL}}{2}=\cosh(\beta mL)\leq e^{(mL\beta)^2/2}.
\end{align*}
The proof is completed.
\end{proof}


\begin{t6}[Remark (\ref{remark 2})]
Consider
\begin{align}\label{consistency:eq11}
 \mathbb{E}(Y^\circ-\hat Y)^2=
& E\bigg(
(X'(\beta^\circ-\hat\beta^{})+H_{(p)}(\phi^\circ-\hat\phi^{})+H_{(q)}(\theta^\circ-\hat\theta^{}))' \nonumber\\
& \hspace{40pt} (X'(\beta^\circ-\hat\beta^{})+H_{(p)}(\phi^\circ-\hat\phi^{})+H_{(q)}(\theta^\circ-\hat\theta^{}))
\bigg).
\end{align}
On the other hand
\begin{align}\label{consistency:eq12}
&\frac{1}{n}\Vert Y^\circ-\hat Y \Vert^2=  \frac{1}{n}(X'(\beta^\circ-\hat\beta^{})+H_{(p)}(\phi^\circ-\hat\phi^{})+H_{(q)}(\theta^\circ-\hat\theta^{}))' \nonumber\\
&\hspace{110pt}(X'(\beta^\circ-\hat\beta^{})+H_{(p)}(\phi^\circ-\hat\phi^{})+H_{(q)}(\theta^\circ-\hat\theta^{})).
\end{align}
Combining (\ref{consistency:eq11}) and (\ref{consistency:eq12}) results in,
{\small
\begin{equation}\label{MSPE equaion:1}
\begin{aligned}
 \mathbb{E}(Y^\circ-\hat Y)^2-\frac{1}{n}\Vert Y^\circ-\hat Y \Vert^2 & \leq
  4K_\lambda^2 \max_{(1 \leq j \leq r,1 \leq k \leq r)}|\mathbb{E}(X_{j}X'_{k})-\frac{1}{n}X_{j}X'_{k}|\\
 & +4 K_\gamma^2 \max_{(1 \leq j \leq p,1 \leq k \leq p)}|\mathbb{E}(H'_{{(p)}_{j}}H{_{{(p)}_{k}}})-\frac{1}{n}H'_{{(p)}_{j}}H{_{{(p)}_{k}}}| \\
 & +4 K_\tau^2 \max_{(1 \leq j \leq q,1 \leq k \leq q)}|\mathbb{E}(H'_{{(q)}_{j}}H{_{{(q)}_{k}}})-\frac{1}{n}H'_{{(q)}_{j}}H{_{{(q)}_{k}}}| \\
 & +8K_\lambda K_\gamma \max_{(1 \leq j \leq r,1 \leq k \leq p)}|\mathbb{E}(X_{{}_{j}}H{_{{(p)}_{k}}})-\frac{1}{n}X_{{}_{j}}H{_{{(p)}_{k}}}|\\
 & + 8 K_\gamma K_\tau \max_{(1 \leq j \leq p,1 \leq k \leq q)}|\mathbb{E}(H'_{{(q)}_{j}}H{_{{(p)}_{k}}})-\frac{1}{n}H'_{{(q)}_{j}}H{_{{(p)}_{k}}}| \\
 & + 8K_\lambda K_\tau \max_{(1 \leq j \leq r,1 \leq k \leq q)}|\mathbb{E}(X_{{}_{j}}H{_{{(q)}_{k}}})-\frac{1}{n}X_{{}_{j}}H{_{{(q)}_{k}}}|.
\end{aligned}
\end{equation}
}
\normalsize
Take $K^*=max\{K_\lambda^2, K_\gamma^2, K_\tau^2,2K_\lambda K_\gamma,2K_\lambda K_\tau,2 K_\gamma K_\tau\}$ and notes that each term in the max expressions is bounded (e.g. $|\mathbb{E}(X_{j}X'_{k})-X_{ij}X'_{ki}|\leq 2 M_1^2$). Using Lemma  (\ref{lemma2}) and lemma 3 of \cite{sourvan2013}, each term in (\ref{MSPE equaion:1}) is bounded by $2M_iM_j\sqrt{\frac{2log(2 a_ia_j)}{n}}, $
$i \in \{1,2,3, \ldots,r\}$,
$j \in \{1,2,3, \ldots,p\}$,
$k \in \{1,2,3,\ldots,q\}$,
$a_1=r,a_2=p \text{ and }a_3=q$. Then,
\fontsize{10pt}{10pt}
\begin{align*}
 MSPE(\hat\beta^{},\hat\phi^{},\hat\theta^{})
 \leq
 \frac{ 2K_{max} M_{max}\sigma}{\sqrt{n}}
 \sum_{i=1}^{3}
 \left(\sqrt{2\log(2a_i)}
\right)+
8K^*\sum_{i,j=1}^{3}
\left(
M_i M_j\sqrt{\frac{2log(2 a_ia_j)}{n}}
\right).
\end{align*}
\normalsize
The proof is completed.
\end{t6}
\bibliographystyle{plain}
\bibliography{maina}
\end{document}